\documentclass[preprint]{aastex62}
\usepackage{natbib}
\accepted{15 November  2019}                                                                                                                                  

\submitjournal{ApJ}

\begin{document}

\title{X-RAY EMISSION AND DISK IRRADIATION  OF HL TAU AND  
       HD 100546}

\correspondingauthor{Stephen L. Skinner}
\email{stephen.skinner@colorado.edu}

\author{Stephen L. Skinner}
\affiliation{Center for Astrophysics and
Space Astronomy (CASA), Univ. of Colorado,
Boulder, CO, USA 80309-0389}

\author{Manuel  G\"{u}del}
\affiliation{Dept. of Astrophysics, Univ. of Vienna,
T\"{u}rkenschanzstr. 17,  A-1180 Vienna, Austria}

\newcommand{\ltsimeq}{\raisebox{-0.6ex}{$\,\stackrel{\raisebox{-.2ex}%
{$\textstyle<$}}{\sim}\,$}}
\newcommand{\gtsimeq}{\raisebox{-0.6ex}{$\,\stackrel{\raisebox{-.2ex}%
{$\textstyle>$}}{\sim}\,$}}
\begin{abstract}
\small{We present new X-ray observations of the optically-obscured 
protostar HL Tau and the intermediate mass Herbig Be star HD 100546.
Both objects are surrounded by spectacular disks showing complex morphology 
including rings and gaps that may have been sculpted by protoplanets. 
HL Tau was detected as a variable  hard X-ray source by {\em Chandra},
typical of late-type magnetically-active coronal sources. No extended
X-ray emission was seen along the HL Tau jet, or along the jet of
the T Tauri binary system XZ Tau located 23$''$ to its east.
In contrast, HD 100546 was detected by {\em XMM-Newton} as a soft X-ray 
source (kT $\ltsimeq$ 1 keV) with  no short-term ($<$1 day) variability. 
Its X-ray properties  are remarkably  similar to the Herbig stars AB Aur 
and HD 163296, strongly suggesting that their X-ray emission arises from
the same mechanism and is intrinsic to the Herbig stars themselves, not
due to unseen late-type companions. We consider several possible emission 
mechanisms and conclude that the X-ray properties of HD 100546 
are consistent with an accretion shock origin, but higher resolution grating 
spectra capable of providing information on individual emission lines are needed to more 
reliably distinguish between accretion shocks and alternatives.
We show that X-ray ionization and heating are mainly confined to the upper disk 
layers in both HL Tau and HD 100546, and any exoplanets near the midplane at
distances $\geq$1 au are well-shielded from X-rays produced by the central star.
}
\end{abstract}
\section{Introduction}
Observational studies of pre-main sequence (PMS) stars and their protoplanetary 
disks provide essential information on the environment in which exoplanets 
form that is needed to test different planet formation models. Young stars of
ages a few Myr that are still accreting material exhibit a diverse range of phenomena 
which can affect disk physical and chemical properties, disk lifetimes, and planet 
formation and evolution. These include excess UV emission associated with accretion shocks, 
intense X-ray emission due to strong magnetic activity or shocks, and angular 
momentum dispersal via wide angle winds or collimated jets and ouflows. 
A review of these processes and their potential effects on protoplanetary disks
and planet formation can be found in Alexander et al. (2014).

The X-ray luminosity of low and intermediate mass PMS stars is typically 
in the range  L$_{x}$ $\sim$ 10$^{28}$ - 10$^{31}$ ergs s$^{-1}$, about
10$^1$-10$^{4}$ times that of the quiet Sun. As  young solar-like  
stars age and spin down, their X-ray luminosity declines 
(G\"{u}del, Guinan, \& Skinner 1997). Stellar X-rays
heat and ionize disk gas and the gaseous atmospheres of any
close-in protoplanets, and must be taken into account in realistic
models of planet formation. X-ray ionization of disk gas also promotes 
the magneto-rotational instability (Balbus \& Hawley 1991) and can thereby 
affect accretion onto the central star.
In magnetically-active class I protostars obscured by surrounding envelopes
and optically-revealed T Tauri stars (TTS), very high plasma temperatures (T $\sim$ 100 MK) 
can be reached during powerful short-duration (minutes to hours) magnetic 
reconnection flares. In addition to flash-heating, such flares can produce 
high energy particles that bombard the inner disk and any close-in 
protoplanets (reviewed by Feigelson \& Montmerle 1999; Feigelson 2010).

The photoelectric cross section for absorption of X-rays by gas
scales inversely with photon energy as
$\sigma$ $\propto$ E$^{-p}$, where $p$ $\approx$ 2.5 for 
solar-abundance gas.  Soft X-rays (E $\ltsimeq$ 1 keV) are
thus heavily absorbed in outer disk layers, whereas harder
X-rays with energies of several keV penetrate deeper and 
can potentially reach the disk midplane, as can cosmic rays.
As a young star evolves and its inner disk gas disperses, 
close-in protoplanets become more heavily exposed to stellar X-rays.
Detailed discussions of X-ray heating and
ionization of disk gas have been presented by
Glassgold et al. (1997a, 1997b. 2004), Igea \& Glassgold (1999),
Shang et al. (2002), and Aresu et al. (2012).
The effects of X-ray and EUV heating on 
exoplanet atmospheres have been studied by numerous authors
including Lammer et al. (2003), Cecchi-Pestellini et al. (2006), 
and Owen \& Jackson (2012).

Compared to the large number of main-sequence exoplanet host 
stars which have now been identified, there are relatively few
examples of PMS exoplanet hosts. One of the first PMS stars
showing evidence for one or more protoplanets was the 
TTS LkCa 15 (Kraus \& Ireland 2012; Sallum et al. 2015). 
It is a bright X-ray source 
(L$_{x}$=10$^{30.5}$ ergs s$^{-1}$), but X-ray heating
and ionization are largely confined to the outer disk layers
and the innermost candidate protoplanet at $r$ $\approx$ 16-20 au
is still well-shielded by disk gas (Skinner \& G\"{u}del 2017, hereafter SG17).    
More recently, the detection of a young protoplanet within
the inner disk gap of the TTS PDS 70 has been reported (Keppler et al. 2018).

Of primary interest here are two other nearby PMS stars which may host
protoplanets, namely the heavily-obscured low-mass protostar HL Tau 
and the intermediate-mass Herbig Be (HBe)
star HD 100546. Their properties are summarized in Table 1 and discussed
further below (Sec. 2). Both objects are surrounded by spectacular well-studied
disks showing complex morphology including rings and gaps that may have been
sculpted by protoplanets. However, the confirmed detection of protoplanets 
is still lacking for both stars and more direct observational evidence is needed.

Both stars are X-ray sources and their proximity ($\approx$110 - 140 pc)
makes them exemplary targets for quantifying the X-ray properties of 
PMS stars which may host exoplanets.
In addition, the earlier B9Ve spectral type and higher mass of HD 100546 
provide an informative X-ray comparison with the solar-mass protostar HL Tau.
Class I protostars and T Tauri stars have convective envelopes capable of sustaining magnetic
fields. They show multiple signatures of magnetic activity including hard
and often variable X-ray emission (including powerful flares), analogous to coronal X-ray sources
like the Sun. Magnetic fields are also thought to control accretion in
these low mass young stellar objects by channeling infalling gas along 
field lines (Hartmann et al. 2016). 

In contrast, Herbig stars are more massive ($\gtsimeq$2 M$_{\odot}$) and have higher effective
temperatures (T$_{eff}$ $\gtsimeq$ 8000 K). Although X-ray emission is
commonly detected in both Herbig Ae and Be stars, its origin is not 
yet understood and its effect on their circumstellar disks has not
been thoroughly studied. It is believed that in some cases the X-rays may 
originate in unseen late-type companions (Skinner et al. 2004; Stelzer et al. 2006).  
However, as we show here, some Herbig stars such as AB Aur, HD 163296, and
HD 100546 have remarkably similar X-ray luminosities and soft X-ray
emission (kT $\ltsimeq$ 1 keV) that suggest intrinsic emission from the 
Herbig stars themselves, not late-type companions. A key unresolved 
question is what role, if any, magnetic fields play in Herbig star X-ray
production. A solar-like dynamo with associated convection and
magnetic activity is not anticipated in  Herbig stars 
with radiative atmospheres. Nevertheless, young Herbig stars of ages 
$\ltsimeq$ a few Myr may still possess weak magnetic fields inherited 
from the progenitor molecular cloud and nonsolar mechanisms have also
been proposed by which Herbig stars might produce magnetic fields.
These mechanisms, along with other potential X-ray processes such as
accretion shocks and shocked winds are discussed further in Sec. 5.2.

Comparisons of low and intermediate-mass PMS stars provide insight
into similarities and differences in their planet-forming environments.
Several factors can lead to harsher environments around intermediate-mass 
PMS stars including their higher effective temperatures, more intense
stellar radiation fields (especially in the UV), and stronger winds
(Stahler \& Palla 2004).
We present here a comparison of the X-ray properties of
the class I protostar HL Tau and the Herbig Be star HD 100546 
based on new  pointed observations
with the {\em Chandra X-ray Observatory} (CXO) and the {\em XMM-Newton}
observatory, respectively (Table 2). Archive data are also discussed. Significant
X-ray differences are found, with HL Tau showing harder, variable,  and more 
heavily absorbed emission. We discuss the implications of these
differences for X-ray production mechanisms and for X-ray irradiation of the disk.

We also take advantage of {\em Chandra}'s arcsecond angular resolution to search for 
X-ray emission along  the HL Tau jet at small offsets from the star,
as well that of its jet-driving TTS neighbor XZ Tau. Faint soft X-ray emission
extending outward a few arcseconds along the optically-traced jets of a 
few PMS stars has been reported. These X-ray jet detections include
DG Tau (G\"{u}del et al.2008), RY Tau (Skinner et al. 2016), 
L1551 IRS 5 (Bally et al. 2003), and the close binary Z CMa (Stelzer et al. 2009). 
Since the sample of young stars with detected X-ray jets is small and the 
mechanism(s) by which the jets are heated to X-ray temperatures 
of a few MK are not understood, further searches are needed
to clarify the ubiquity and origin of this unusual phenomenon. 
{\em Chandra}'s low detector background and high spatial resolution 
make it ideally suited for X-ray jet searches.

\newpage

\section{Summary of Stellar properties}

\vspace*{0.1cm}

\subsection{HL Tau} 

HL Tau is an optically-obscured young star in the Taurus
star-forming region which was classified as a class I protostar
on the basis of its mid-IR spectral index by White \& Hillenbrand (2004).
The central star is not seen directly at optical 
wavelengths but is revealed in scattered light,
so the stellar properties (Table 1) are somewhat uncertain. 
Stellar mass estimates suggest  
M$_{*}$ $\approx$ 1.2 - 1.3 M$_{\odot}$ 
(Brogan et al. 2015; Robitaille et al. 2007; White \& Hillenbrand 2004).
Estimates of the accretion rate span a  range
from log $\dot{M}_{acc}$ = $-$7.06 to $-$5.05 M$_{\odot}$ yr$^{-1}$
(Hayashi et al. 1993; White \& Hillenbrand 2004; Beck et al. 2010).
 
Infalling gas from an extended disk-shaped envelope was
detected in $^{13}$CO observations by Hayashi et al. (1993).
High spatial resolution Atacama Large Millimeter Array (ALMA) 
mm and sub-mm observations
analyzed by the ALMA Partnership (Brogan et al. 2015)
revealed dramatic disk structure dominated by a pattern
of bright and dark rings. The ring gaps are dust-depleted 
(Pinte et al. 2016) and the presence of orbital resonances 
suggests that the rings may have been sculpted by planet formation. 
Molecular line velocities consistent with Keplerian disk 
motion around a $\sim$1.3 M$_{\odot}$ central star were
detected by ALMA, in good agreement with the previous mass 
estimates referenced above.

HL Tau drives a bipolar jet with several emission knots revealed 
in optical images (Mundt et al. 1990; Movsessian et al. 2012). 
Knots are traced in H$\alpha$ outward to at least 35$''$ along the 
blueshifted jet axis at P.A. $\approx$ 45$^{\circ}$ - 47$^{\circ}$.
The inner jet is visible in H$\alpha$ a few arcseconds from the
star. The jet includes low and high
radial velocity structures and the mean absolute jet velocity
is $v_{jet}$ $\approx$ 250 km s$^{-1}$ (Movsessian et al. 2012).
Interestingly, the blueshifted jet bends at a separation of 
20$''$ from the star, possibly due to the jet's interaction  
with matter expelled by the wind of the young binary XZ Tau
located 23$''$ to the east (Movsessian et al. 2012).
We discuss the X-ray properties of HL Tau and XZ Tau and describe 
our search for X-ray emission from their jets in Sections 4.1 and 4.2.

\subsection{HD 100546}

HD 100546 is a nearby Herbig Be star that has been 
extensively studied in the UV, optical, IR, 
and mm, as summarized by Sissa et al. (2018).
Stellar parameters (Table 1) have been updated based on the recent
{\em Gaia} DR2 distance of 110 pc by Pineda et al. (2019)
and Vioque et al. (2018). Interstellar  extinction is low 
with most studies giving a range A$_{\rm v}$ = 0.1 - 0.28 mag 
(e.g. van den Ancker et al. 1998; Deleuil et al. 2004; Pineda et al. 2019).
However, larger values A$_{\rm v}$ $\approx$ 1 mag that probably 
include circumstellar extinction have been deduced from 
fitting the continuum spectral energy distribution over
a broad wavelength range (Elia et al. 2004).

The IR spectrum of HD 100546 bears a remarkable resemblance
to comet Hale-Bopp (Malfait et al. 1998), suggesting that
the star is surrounded by a huge comet cloud (Grady et al. 1997).
The HD 100546 circumstellar disk has complex morphology
including a gap (Jamialahmadi et al. 2018; Sissa et al. 2018),
spiral-shaped structures (Grady et al. 2001, 2005;
Ardila et al. 2007), and rings (Sissa et al. 2018).
Evidence for accreting gas based on UV spectral line
profiles has been reported (Grady et al. 1997; Deleuil et al.  2004).
P Cygni-type line absorption profiles provide solid evidence
for a high-velocity wind (Grady et al. 2005). 

Very Large Telescope (VLT) adaptive optics  mid-IR 
$L'$ and $M'$ images revealed a source
at a separation of 0$''$.457 - 0$''$.48 ($\sim$50 - 53 au)
from the star that was
classified as a candidate giant protoplanet
(HD 100546b; Quanz et al. 2013; 2015). This object
was subseqently detected in $H$-band Gemini Planet
Imager (GPI) images by Currie et al. (2015), but
they noted that it could be a locally bright region
in the disk and not a protoplanet.
The  GPI observations
of Rameau et al. (2017) detected no significant
orbital motion of this source on a timescale of 4.6 years
and found an H-band spectrum inconsistent with a
low-temperature object, suggesting that the source
may be scattered light. No H$\alpha$ emission was detected
from the candidate protoplanet using VLT data (Cugno et el. 2019)
nor was any 870 $\mu$m continuum emission detected by
ALMA (Pineda et al. 2019). Thus, definitive confirmation
of HD 100546b as a protoplanet is yet to be obtained.

A second candidate protoplanet referred to as HD 100546c
located closer to the star
at a projected separation of $\sim$10-14 au was inferred
by Brittain et al. (2015) and Currie et al. (2015).
However,  neither H$\alpha$ emission (Cugno et el. 2019)  
nor compact 870 $\mu$m emission (Pineda et al. 2019)
was detected from HD 100546c.
Furthermore, its presence was not confirmed in the high-constrast
near-IR study of Sissa et al (2018). Thus, the reality
of HD 100546c as a protoplanet remains questionable.

\begin{deluxetable}{lllccccccl}
\tablewidth{0pt}
\tablecaption{Stellar Properties}
\tablehead{
           \colhead{Name}               &
           \colhead{Sp. Type}           &
           \colhead{Age}                &
           \colhead{M$_{*}$}            &
           \colhead{R$_{*}$}            &           
           \colhead{T$_{eff}$}          &
           \colhead{L$_{*}$}            &
           \colhead{A$_{\rm V}$}        &
           \colhead{distance}           &
           \colhead{Refs.\tablenotemark{a}}   \\
           \colhead{}                   &
           \colhead{}                   &
           \colhead{(Myr)}              &
           \colhead{(M$_{\odot}$)}      &
           \colhead{(R$_{\odot}$)}      &
           \colhead{(K)}                &
           \colhead{(L$_{\odot}$)}      &
           \colhead{(mag)}              &
           \colhead{(pc)}               &
           \colhead{}                   
                                  }
\startdata
HL Tau    & K5$\pm$1 & $\leq$2       & 1.2-1.3               & 2.1\tablenotemark{b} & 4400 & 1.5-3.0\tablenotemark{c}    & 7.4   & 140 & 1,2,3,4     \\
HD 100546 & B9Vne    & $\sim$4.8-5.5 & 2.2\tablenotemark{d}  & 1.7\tablenotemark{b} & 9750 & 24                          & 0.1-1 & 110 & 5,6,7,8,9,10 \\
\enddata
\tablenotetext{a}{
References: (1) White \& Hillenbrand (2004) (2) Robitaille et al. (2007) (3) Beck et al. (2010) 
(4) Brogan et al. (2015
(5) van den Ancker et al. (1998) (6) Pineda et al. (2019) (7) Fairlamb et al. (2015)
(8) Vioque et al. (2018) (9) Elia et al. (2004) (10) {\em Gaia} DR2 distance  
}
\tablenotetext{b}{R$_{*}$ is the equivalent blackbody radius computed using
L$_{*}$ = 1.53 L$_{\odot}$, T$_{eff}$ = 4395 K for HL Tau (White \& Hillenbrand 2004) and
L$_{*}$ = 23.4 L$_{\odot}$, T$_{eff}$ = 9750 K for HD 100546 (Vioque et al. 2018).}
\tablenotetext{c}{L$_{bol}$ = 6.6 L$_{\odot}$ (White \& Hillenbrand 2004).}
\tablenotetext{d}{Recent mass estimates based on the {\em Gaia} DR2 distance (110 pc) are
                  M$_{*}$ = 2.05 M$_{\odot}$ (Vioque et al. 2018) and 
                  M$_{*}$ = 2.2$\pm$0.2 M$_{\odot}$ (Pineda et al. 2019).}
\end{deluxetable}

\section{X-ray Observations} 

\subsection{HL Tau }

{\em Previous Observations}:~
HL Tau  has been serendipitously captured off-axis in a few previous
{\em Chandra} observations. It was located closest to the aimpoint 
(but still 6$'$.15 off-axis) in a 65 ks ACIS-S observation of HH 154 
in December 2009 (ObsId 11016). Because of the off-axis position,
the image is blurred and not suitable for determining whether
faint extended jet emission close to the star is present. 
However, we extracted a spectrum and X-ray light curve  using tools
provided in the  Chandra Interactive Analysis of Observations (CIAO) 
software package, as described below for the new observations.
The spectrum was acceptably fitted with a simple one-temperature 
(1T) APEC thermal plasma model. Fit results are included in
Table 3 for comparison with the new observations.
The spectrum shows very little flux below 1 keV. A faint Fe K complex 
(Fe XXV; E = 6.67 keV) emission line  is detected, 
indicative of very hot plasma. 
No large-amplitude flares were detected but low-level variability may 
be present in the X-ray light curve. However, a large X-ray flare
which reached temperatures of kT $\approx$ 7-8 kev and  decayed over a 
two-day interval was detected in the 5-day {\em XMM-Newton} monitoring 
program discussed by Giardino et al. (2006). Excluding the large flare,
they determined the quiescent spectral parameters to be absorption 
N$_{\rm H}$ = 2.43 $\times$ 10$^{22}$ cm$^{-2}$ and plasma temperature
kT = 3.1 keV.
HL Tau was also brightly 
detected in the {\em XMM-Newton} Extended Survey of the Taurus
Molecular Cloud (XEST), as discussed by G\"{u}del et al. (2007).
Spectral parameters given in the XEST catalog based on two-temperature
thermal plasma fits are 
N$_{\rm H}$ = 2.79 (2.55-3.21; 1$\sigma$) $\times$ 10$^{22}$ cm$^{-2}$,
kT$_{1}$ = 1.93 keV, kT$_{2}$ = 13.8 keV, emission-measure 
weighted temperature kT$_{wgtd}$ = 3.0 keV, and 
intrinsic (unabsorbed) X-ray luminosity log L$_{x}$(0.3-10 keV) = 30.52 ergs s$^{-1}$.

{\em New Observations}:~
As summarized in Table 2,
we observed HL Tau in two {\em Chandra} exposures acquired in 
Dec. 2017 (ObsId 20906) and Jan. 2018 (ObsId 18915) using the 
Advanced CCD Imaging Spectrometer (ACIS-S).
In contrast to the previous {\em Chandra} observations, HL Tau was
positioned on-axis to minimize the point-spread-function (PSF) 
broadening. This allowed us to perform image deconvolution to 
remove PSF blurring and search for faint X-ray emission from
the jet at small offsets within a few arcseconds of the star. 
ACIS-S was configured in  1/4 subarray mode using a short 
0.9 s CCD frame time to mitigate any photon pileup in the event of a 
large X-ray flare. No large flares occurred and pileup was 
negligible ($\leq$2\%).  

Data were reduced using CIAO version 4.11 in combination with
recent calibration data (CALDB vers. 4.8.2).
X-ray spectra and associated response matrix files (RMF) and 
auxiliary response files (ARF) files were extracted 
using CIAO {\em specextract}. Energy-filtered light curves
were produced using CIAO {\em dmextract}. 
Spectra and light curves were extracted from a circular region
of radius 1$''$.5 centered on the source for the new on-axis 
observations, but a larger radius of 7$''$ was used for the 
off-axis archive observation to fully enclose the blurred source. 
Background was negligible and no time filtering to remove high background
intervals was required. Spectra were analyzed using 
XSPEC vers. 12.10.1 (Sec. 4.1; Table 3). Image deconvolution
was performed using CIAO {\em arestore}, which is based on
the Lucy-Richardson method.

\subsection{HD 100546}

{\em Previous Observations}:
HD 100546  was detected as an X-ray source in two short 2.6 ks
{\em Chandra} observations obtained in 2002 (Feigelson et al. 2003).
Their fit of the spectrum (59 cts, 0.5-8 keV) gave a plasma 
temperature kT = 2.5 keV and intrinsic X-ray luminosity 
log L$_{x}$(0.5-8 keV) = 29.46 ergs s$^{-1}$, where we have
adjusted their published L$_{x}$ value slightly upward ($+$0.06 dex)
using the more recent {\em Gaia} DR2 distance of 110 pc.

{\em New Observation}:
We observed HD 100546 in a single {\em XMM-Newton} observation
of duration $\approx$76 ks in July 2015 (Table 2).
The primary instrument was the
European Photon Imaging Camera (EPIC) which provides
CCD imaging spectroscopy from the pn camera (Str\"{u}der et al. 2001)
and two nearly identical MOS cameras (MOS1 and  MOS2; Turner et al. 2001).

Data were reduced using the {\em XMM-Newton} Science Analysis System
(SAS vers. 17.0) with recent calibration data. Event files provided
by the {\em XMM-Newton} Science Operations Center were filtered to select
good event patterns. Time filtering was applied to the pn data to
remove intervals of high background radiation, resulting in 66.3 ks of good
time interval (GTI) pn exposure and 58.98 ks of pn livetime.
No time filtering was required for the MOS cameras, which
are less affected by high background radiation. Each MOS camera provided
$\approx$74.7 ks of livetime. Energy filters were
applied to the pn and MOS data for light curve extraction in order
to improve signal-to-noise ratio.

Background-subtracted X-ray spectra and light curves were extracted
for HD 100546 using a circular region centered on the X-ray source of
radius 20$''$.
Background was extracted from source-free regions near the source.
Observation-specific RMF and ARF files were created using SAS tools. 
Spectra were analyzed using XSPEC vers. 12.10.1 \\

\begin{deluxetable}{llll}
\tabletypesize{\small}
\tablewidth{0pt}
\tablecaption{New X-ray Observations  }
\tablehead{
\colhead{Parameter} &
\colhead{} \\
}
\startdata
Star                 & HD 100546              & HL Tau                     & HL Tau     \\
Telescope            & XMM-Newton             & Chandra                    & Chandra    \\
ObsId                & 0761790101             & 20906                      & 18915      \\
Start Date/Time (TT) & 2015 Jul. 10/14:14:43  & 2017 Dec. 27/16:21:32      & 2018 Jan. 6/13:31:51     \\
Stop  Date/Time (TT) & 2015 Jul. 11/10:5024   & 2017 Dec. 28/03:17:50      & 2018 Jan. 6/22:08:20     \\
Instrument           & EPIC\tablenotemark{a}  & ACIS-S\tablenotemark{b}    & ACIS-S\tablenotemark{b}  \\
Livetime (s)\tablenotemark{c}         & 58,980\tablenotemark{d}& 35,994                     & 26,857   \\
Frame time (s)       & 0.073\tablenotemark{d} & 0.9                        & 0.9 \\
\enddata
\tablenotetext{a}{The EPIC observation was obtained in Full Window mode using the thick optical filter.
The energy range is
E $\approx$ 0.2 - 12 keV with energy resolution (FWHM) at 1 keV of $\Delta$$E$ $\approx$
100 eV (pn) and $\Delta$$E$ $\approx$ 70 eV (MOS). 
The spatial resolution at 1.5 keV is 6$''$.6 (FWHM) for pn and 6$''$.0 and 4$''$.5 (FWHM) for
MOS1 and MOS2.
}
\tablenotetext{b}{Data were obtained using ACIS-S 1/4 subarray in faint timed event mode.  
ACIS-S has a pixel size of 0$''$.492 and the 
energy range is E $\approx$ 0.5 - 10 keV.
For an on-axis  point source the 90\% encircled energy radius is 
R$_{90}$ $\approx$ 0$''$.9.
The energy resolution at 1.49 keV is $\approx$130 eV.
}
\tablenotetext{c}{Livetime corresponds to the time during which source data were being collected 
and excludes overhead such as CCD readout times.}
\tablenotetext{d}{Values are for the EPIC pn camera. The pn livetime value of 58.98 ks
takes into account time intervals that were removed due to high background count rate. 
Livetime values for the MOS cameras were 74,646 s (MOS1) and 74,770 s (MOS2).
No high-background time filtering was needed for the MOS data.
}
\end{deluxetable}

\clearpage


\section{Results}

\subsection{HL Tau}

HL Tau was brightly detected in both {\em Chandra} observations,
along with other nearby young stars such as the 
close TTS  binary XZ Tau (Fig. 1-top). 
The peak X-ray position of HL Tau in the two observations agrees
to within one ACIS pixel (0$''$.492). Averaging the peak position for 
the two observations gives a {\em Chandra} J2000 position
R.A. = 04$h$ 31$m$ 38.43$s$, decl. = $+$18$^{\circ}$ 13$'$ 57.4$''$.
For comparison, the ALMA position is (Brogan et al. 2015)
R.A. = 04$h$ 31$m$ 38.425$s$, decl. = $+$18$^{\circ}$ 13$'$ 57.242$''$.
 
Its broad-band X-ray light curve (Fig. 2-top) is clearly variable in ObsId 18915. 
The CIAO $glvary$ statistical test applied to events in the hard 2-8 keV
range gives a probability of variability P$_{var}$ $>$ 0.999, whereas
events in the soft 0.3-2 keV band give P$_{var}$ = 0.20. 
The hard-band variability is apparent in Figure 2-bottom.
By comparsion, the test for variability in ObsId 20906 does not show
significant variability, with $glvary$ giving P$_{var}$ = 0.49 (0.3-2 keV)
and 0.59 (2-8 keV). We generated plots of the hardness ratio (H.R.) 
for both observations showing the ratio of soft to hard-band count rates
versus time. When binned at 1800 s intervals, a fit of the ObsId 18915
H.R. plot with a constant rate model gives a mean H.R = 0.30$\pm$0.02 (1$\sigma$)
and ObsId 20906 gives H.R = 0.39$\pm$0.03 (1$\sigma$). This provides quantitative
confirmation of the lower ratio of soft to hard band rates apparent 
for ObsId 18915 in Fig. 2-bottom.

Deconvolution was performed on the ObsId 20906
soft-band (0.3-2 keV) image to search for extended 
soft X-ray emission along the optical jet. The deconvolution
was performed using a PSF image generated using the
MARX simulation software based on parameters 
specific to ObsId 20906. No significant X-ray extension in the
optical jet direction was found (Fig. 1-bottom).
Deconvolution of the ObsId 18915 image was not performed
since the spectrum and hardness were highly variable.
Such variability causes the energy-dependent PSF to 
change shape during the observation.

The ACIS-S spectra (Fig. 3-left) reveal hot plasma
including a faint Fe K  (Fe XXV) blended emission
line complex at 6.67 keV in ObsId 18915.
Spectral fits using an absorbed single-temperature optically
thin plasma model are acceptable (Table 3). The best-fit
values in Table 3 for ObsId 18915 should be interpreted
as time averages, since the spectral parameters were
changing during the exposure (see below).
The fits require substantial aborption equivalent
to A$_{\rm V}$ $\approx$ 13 mag, using the conversion
N$_{\rm H}$ = 1.9 $\times$ 10$^{21}$A$_{\rm V}$ cm$^{-2}$
obtained by averaging the results of Gorenstein (1975)
and Vuong et al. (2003). Similarly, XMM-Newton {\em XEST}
results (G\"{u}del et al. 2007) give 
N$_{\rm H}$ = 2.79 $\times$ 10$^{22}$ cm$^{-2}$, 
or A$_{\rm V}$ $\approx$ 14.7 mag.

Since the X-ray emission was clearly variable during ObsId 18915,
we subdivided the full  exposure into three equal time segments
of $\approx$9.4 ks each and extracted separate spectra for each segment.
Each time-partitioned spectrum was fitted separately using a
using a 1T APEC model with absorption and metallicity held fixed at the
best-fit values obtained by fitting the entire observation (Table 3).
The simple 1T model gives a poor fit for the first segment, 
suggesting that plasma properties (e.g. temperature) were changing  
rapidly during the first 9.4 ks.
Fits of the middle and last segments are acceptable.
The 1T APEC fit of the middle segment when the hard-band count rate 
was highest gives kT = 4.3 (3.9-5.0; 1$\sigma$) keV and observed
(absorbed) flux 
F$_{x,abs}$(0.3-8 keV) = 1.04 $\times$ 10$^{-12}$ ergs cm$^{-2}$ s$^{-1}$.
A similar fit of the final 9.4 ks segment when the hard-band count
rate was decreasing gives kT = 3.1 (2.8-3.6) keV and
F$_{x,abs}$(0.3-8 keV) = 6.4 $\times$ 10$^{-13}$ ergs cm$^{-2}$ s$^{-1}$.
Thus, the plasma temperature clearly decreased during the final
third of the observation. Even so, it had not yet decreased to
the lower levels observed about nine days prior in ObsId 20906 (Table 3).

\subsection{XZ Tau}

We briefly summarize the X-ray properties of the multiple system
XZ Tau AB, even though it is so far not known to host exoplanets.
It consists of an optically variable close $\approx$M2$+$M3 
pair  separated by $\approx$0$''$.3, both of which have characteristics of 
classical TTS  (Haas et al. 1990; White \& Ghez 2001; 
Hartigan \& Kenyon 2003; Krist et al. 2008).
It was suggested from {\em Very Large Array} observations
that the southern component XZ Tau A may itself be a tight binary with
a separation of 0$''$.09 (Carrasco-Gonz\'{a}lez et al. 2009) but 
confirmation of this putative third component has so far not been obtained.
{\em Hubble Space Telescope} observations show that both A and B
drive collimated jets and reveal bipolar expanding bubbles of
emission nebulosity. The XZ Tau A jet is oriented along
P.A. = 15$^{\circ}$ and the B jet along P.A. = 36$^{\circ}$ (Krist et al. 2008).

X-ray emission from the unresolved XZ Tau AB pair was detected during the  
{\em XMM-Newton} monitoring campaign discussed by Giardino et al. (2006)
with average spectral parameters N$_{\rm H}$ = 2.4$\pm$0.1  $\times$ 10$^{21}$ cm$^{-2}$,
kT$_{1}$ = 0.83$\pm$0.02 keV,  kT$_{2}$ = 4.09$\pm$0.34 keV, and
log L$_{x}$ = 30.36 ergs s$^{-1}$. Analysis of {\em XMM-Newton} XEST data by
G\"{u}del et al. (2007) gave
N$_{\rm H}$ = 2.4 (2.1-2.7) $\times$ 10$^{21}$ cm$^{-2}$,
kT$_{1}$ = 0.75 keV, kT$_{2}$ = 2.3 keV, kT$_{wgtd}$ = 1.5 keV, and
log L$_{x}$(0.3-10 keV) = 29.93 ergs s$^{-1}$.

XZ Tau AB was  detected as a bright source in our {\em Chandra} observations.
The close pair is not spatially resolved by {\em Chandra}.
Pileup was negligible ($\leq$2\%) due to the short subarray readout time. 
No significant X-ray variability
was found during either observation. The CIAO {\em glvary} test using
events in the 0.3 - 8 keV range gives variability probabilities
P$_{var}$ $<$ 0.01 (ObsId 18915) and P$_{var}$ = 0.04 (ObsId 20906).
However, the X-ray emission brightened somewhat during the
interval between the observations (see below).
We constructed a deconvolved 0.3-2 keV image (ObsId 20906) using the 
same procedure as for HL Tau. No significant X-ray extension was
found at offsets $\geq$1$''$ along the jet axes of XZ Tau A 
at P.A. = 15$^{\circ}$ or  XZ Tau B at P.A. = 36$^{\circ}$ (Krist et al. 2008).

The spectra for the two observations (Fig. 4) are similar and show
emission lines from Mg XII Ly$\alpha$ (1.47 keV, maximum line emissivity
at log T$_{max}$ =  7.0 K), the Si XIII triplet (1.86 keV, log T$_{max}$ =  7.0 K),
and S XV (2.43 keV, log T$_{max}$ =  7.1 K). The spectra can be 
satisfactorily fitted with a two-temperature (2T) APEC thermal
plasma model and results for fitting the two spectra together are
given in Table 3. The absorption column density
is low and not well-constrained by the {\em Chandra} spectra.
Thus, it was held fixed during fitting at 
N$_{\rm H}$ = 2 $\times$ 10$^{21}$ cm$^{-2}$, compatible
with A$_{\rm V}$ = 1.1 ($+$0.4,$-$0.2) mag for
XZ Tau A (Cs\'{e}p\'{a}ny et al. 2017) and similar to
values determined from previous {\em XMM-Newton} observations.

Even though no significant X-ray variability was present in the 
individual observations, a comparison of the spectra for the
two observations shows that the source increased in brightness.
For the first observation
(ObsId 20906), XZ Tau AB provided 1531 net counts (0.3-8 keV)
during a livetime of 35,994 s (42.5 c ks$^{-1}$) and an
observed (absorbed) flux
F$_{x}$(0.3-8 keV) = 
3.29 (3.10-3.34; 1$\sigma$) $\times$ 10$^{-13}$ ergs cm$^{-2}$ s$^{-1}$.
For the second observation (ObsId 18915), the source
gave 1424 net counts in a livetime of 26,857 s (53.0 c ks$^{-1}$) 
and an observed flux
F$_{x}$(0.3-8 keV) = 
4.08 (3.75-4.13; 1$\sigma$) $\times$ 10$^{-13}$ ergs cm$^{-2}$ s$^{-1}$.
Thus, the XZ Tau X-ray flux is variable. In addition, the somewhat lower
temperature measured by {\em Chandra} as compared to the previous  
{\em XMM-Newton} values shows that kT is variable, as was also 
apparent in the {\em XMM-Newton} monitoring data presented by 
Giardino et al. (2006).

\begin{deluxetable}{lllll}
\tabletypesize{\scriptsize}
\tablewidth{0pc}
\tablecaption{{\em Chandra} X-ray Spectral Fits for HL Tau and XZ Tau}
\tablehead{
\colhead{Parameter}      &
\colhead{       }        &
\colhead{       }        &
\colhead{       }
}
\startdata
Star                                      & HL Tau              & HL Tau               & HL Tau                & XZ Tau    \\
ObsId                                 & 11016\tablenotemark{a}  & 18915                & 20906                 & 18915+20906                 \\
Model                                     & 1T APEC             & 1T APEC              & 1T APEC               & 2T APEC                      \\
N$_{\rm H}$ (10$^{22}$ cm$^{-2}$)         & 2.4 [2.2-2.5]       & 2.5 [2.3-2.7]        & 2.6 [2.4-2.8]         & \{0.2\}                      \\
kT$_{1}$ (keV)                            & 3.04 [2.80-3.31]    & 4.07 [3.56-4.75]     & 2.40 [2.20-2.68]      & 0.26 [0.23-0.30]             \\
kT$_{2}$ (keV)                            & ...                 & ...                  & ...                   & 1.28 [1.24-1.32]             \\
kT$_{wgtd}$ (keV)                         & ...                 & ...                  &                       & 0.78   \\                              
norm$_{1}$ (cm$^{-5}$)                    & 9.9 [9.1-10.8]e-04  & 12.4 [11.0-13.9]e-04 & 9.6 [8.4-11.0]e-04    & 5.7 [4.4-7.3]e-04           \\
norm$_{2}$ (cm$^{-5}$)                    & ...                 & ...                  & ...                   & 5.9 [5.5-6.3]e-04           \\
$Z$/$Z_{\odot}$\tablenotemark{b}          & 0.44 [0.32-0.57]    & 0.43 [0.30-0.58]     & 0.43 [0.24-0.64]      & 0.14 [0.12-0.17]             \\
$\chi^{2}$/dof                            & 177.8/178 & 101.8/107            & 73.9/80               & 201.9/186                  \\
reduced $\chi^{2}$                        & 1.00      & 0.95                 & 0.92                  & 1.08                       \\
F$_{\rm x}$ (10$^{-13}$ ergs cm$^{-2}$ s$^{-1}$)\tablenotemark{c} & 4.91 (13.4) & 7.70 (18.4)  & 3.57 (12.0)      & 3.59 (7.57)                \\
log L$_{\rm x}$ (ergs s$^{-1}$)           & 30.50               & 30.63                & 30.45                 & 30.25                      \\
log (L$_{\rm x}$/L$_{bol}$)\tablenotemark{d}  & $-$3.90         & $-$3.77              & $-$3.95               & $-$3.01                    \\
\enddata
\tablecomments{
Based on  simultaneous fits of background-subtracted ACIS-S spectra 
binned to a minimum of 10 counts per bin using XSPEC v12.10.1. 
For XZ Tau, the spectra from both observations were fitted simultaneously.
Models include an absorption
component (N$_{\rm H}$) modeled in XSPEC with the $wabs$ model and are 
one-temperature variable metallicity thermal plasma (1T APEC).
The tabulated parameters
are absorption column density (N$_{\rm H}$), plasma temperature of each component (kT$_{i}$), 
emission-measure weighted plasma temperature (kT$_{wgtd}$) for 2T fits 
weighted by the $norm$  of each component,
metallicity ($Z$) expressed as a fraction of solar metallicity, and
XSPEC normalization (norm).
For APEC models, the volume emission measure 
is related to the normalization (norm) by
n$_{e}^2$V = 4$\pi$$\times$10$^{14}$d$_{cm}^2$$\cdot$norm,
where n$_{e}$ is electron density, V is the volume of 
X-ray emitting plasma, and d$_{cm}$ is the distance in cm.
Quantities enclosed in curly braces were held fixed during fitting.
Square brackets enclose 1$\sigma$ confidence intervals.
X-ray flux (F$_{\rm X}$) is the  observed (absorbed) value followed
in parentheses by the unabsorbed value in the 0.3 - 8 keV range.
X-ray luminosity (L$_{\rm X}$) is the unabsorbed value in the
0.3 - 8 keV range. A distance of 140 pc is assumed.}
\tablenotetext{a}{Archive data, observed on 29 Dec 2009 with HL Tau located 6$'$.15 off-axis (Sec. 3.1).}
\tablenotetext{b}{Element abundances are relative to the solar values of Anders \& Grevesse (1989).}
\tablenotetext{c}{Measurement uncertainties in observed (absorbed) flux 
                  are $\approx$5\% - 7\% (1$\sigma$).}
\tablenotetext{d}{Assumes L$_{bol}$ = 6.6 L$_{\odot}$ for HL Tau (White \& Hillenbrand 2004)
                  and L$_{bol}$(A$+$B) = 0.48 L$_{\odot}$ for XZ Tau AB (Hartigan \& Kenyon 2003). } 
\end{deluxetable}


\subsection{HD 100546}

HD 100546 was clearly detected with a corrected X-ray
position (J2000) from the pipeline processing of
R.A. = 11$h$ 33$m$ 25.38$s$, decl. = $-$70$^{\circ}$ 11$'$ 41$''$.2.
This position is offset by only 0$'$.31 from the ICRS coordinates of 
HD 100546:  R.A. = 11$h$ 33$m$ 25.44$s$,
decl. = $-$70$^{\circ}$ 11$'$ 41$''$.24.
The offset is well within {\em XMM-Newton} astrometric
uncertainties\footnote{{\em XMM-Newton} calibration data can
be found at http://xmm2.esac.esa.int/docs/documents/CAL-TN-0018.pdf~.},
providing confidence that the X-ray source is HD 100546.
There are no other X-ray sources in the immediate
vicinity of HD 100546 (Fig. 5).

The EPIC light curves of HD 100546 are shown in Figure 6.
The light curves have been energy-filtered to include only
events in the 0.2-4 keV range since there are no significant
spectral counts above 4 keV (Fig. 7).
No large amplitude variability (e.g. flares) were detected,
but a slight increase in count rate during the last half of 
the observation may be present. A $\chi^2$  test for variability
in the lower noise MOS light curves binned to 2400 s 
bins gives P$_{var}$ = 0.61 (MOS1) and P$_{var}$ = 0.90 (MOS2).

The EPIC spectra shown in Figure 7 are quite soft and almost all
detected events have energies below 2 keV. A few emission lines
and line blends are visible, as apparent in the lightly-binned
MOS spectra (Fig. 7-right). These are O VIII Ly$\alpha$ (0.65 keV.
log T$_{max}$ = 6.5 K), the blended Ne IX triplet (0.905-0.922 keV,
log T$_{max}$ = 6.6 K), Ne X Ly$\alpha$ (1.02 keV, log T$_{max}$ = 6.8 K)
and Ne X Ly$\beta$ (1.21 keV, log T$_{max}$ = 6.8 K). These lines all
trace plasma at T $<$ 10 MK, much cooler than  detected in HL Tau. 

Spectral fits with absorbed thermal plasma models are summarized
in Tables 4 and 5. Two-temperature optically thin plasma (2T APEC, 2T VAPEC)
and differential emission measure (DEM) models give similar results. 
These models adequately reproduce the spectra but tend to underestimate
the flux in the 1.21 keV feature visible in both pn and MOS that is likely Ne X. 
The EPIC fits converge to a subsolar Fe abundance and Ne abundance above solar,
but grating spectra are needed to more reliably determine element abundances.

All models that we tested require an
absorption column density N$_{\rm H}$ that is 
larger than that expected based on typical interstellar extinction
estimates A$_{\rm V}$ $\approx$ 0.1 - 0.3 mag toward HD 100546.
This is an indication that the X-rays are absorbed
by dust-depleted circumstellar gas that escapes optical detection. 
The X-ray derived N$_{\rm H}$ = (1.8 - 2.7) $\times$ 10$^{21}$ cm$^{-2}$ 
equates to A$_{\rm V}$ $\approx$ 0.9 - 1.4  mag using standard
N$_{\rm H}$ to A$_{\rm V}$ conversions (Gorenstein 1975; Vuong et al. 2003). 
This value is similar to A$_{\rm V}$ = 1.03 mag determined from
{\em IUE} short wavelength UV spectra (Valenti et al. 2000).
Also, the X-ray derived N$_{\rm H}$ is comparable to 
N$_{\rm H}$ = 3 $\times$ 10$^{21}$ cm$^{-2}$   
determined by Grady et al. (2005) from {\em HST} observations.
This  X-ray absorption in excess of the interstellar value
was not detected in previous {\em Chandra} observations. 
As a result,  EPIC estimates of the intrinsic (unabsorbed) X-ray
luminosity (L$_{x}$) are slightly higher than
previously determined from {\em Chandra} data.
The higher absorption inferred from {\em XMM-Newton} spectra could 
reflect real variability in N$_{\rm H}$, as discerned from {\em HST}
data by Grady et al. (2005). However, the better sensitivity of
{\em XMM-Newton} EPIC at low energies below 1 keV where absorption
effects are important may also contribute to differences with {\em Chandra}.

\begin{deluxetable}{llll}
\tabletypesize{\scriptsize}
\tablewidth{0pc}
\tablecaption{{\em XMM-Newton} X-ray Spectral Fits for HD 100546}
\tablehead{
\colhead{Parameter}      &
\colhead{       }        &
\colhead{       }        &
\colhead{       }
}
\startdata
Model                                                 & 2T APEC                & 2T VAPEC              & DEM                      \\
N$_{\rm H}$ (10$^{21}$ cm$^{-2}$)                     & 2.7 [2.4-3.3]          & 2.2 [1.9-2.5]         & 1.8 [1.6-2.0]             \\
kT$_{1}$ (keV)                                        & 0.22 [0.20-0.23]       & 0.39 [0.37-0.40]      & ...                       \\
kT$_{2}$ (keV)                                        & 1.01 [0.99-1.03]       & 1.57 [1.38-1.76]      & ...                       \\
kT$_{wgtd}$ (keV)                                     & 0.52                   & 0.73                  & ...                       \\
norm$_{1}$ (cm$^{-5}$)                                & 14.8 [11.7-22.9]e-05   & 7.5 [6.3-8.9]e-05     & 0.23 [0.18-0.27]e-05          \\
norm$_{2}$ (cm$^{-5}$)                                & 9.1  [7.9-10.2]e-05    & 3.2 [2.8-3.5]e-05     & ...                       \\
$Z$/$Z_{\odot}$\tablenotemark{a}                      & 0.41 [0.34-0.52]       & ...                   &  ...                      \\
Ne\tablenotemark{a}                                   & ...                    & 2.8 [2.5-3.2]         & 3.0 [2.7-3.4]              \\
Fe\tablenotemark{a}                                   & ...                    & 0.19 [0.16-0.23]      & 0.30 [0.24-0.36]           \\
$\chi^{2}$/dof                                        & 280.2/213              & 253.9/212             & 280.2/209                  \\
reduced $\chi^{2}$                                    & 1.31                   & 1.20                  & 1.34                       \\
F$_{\rm x}$ (10$^{-14}$ ergs cm$^{-2}$ s$^{-1}$)\tablenotemark{b}  & 8.34 (30.7)  & 8.46 (20.7)        & 8.58 (17.8)                \\
F$_{\rm x,1}$ (10$^{-14}$ ergs cm$^{-2}$ s$^{-1}$)    & 2.05 (16.6)            & 5.73 (16.0)           & ...                         \\
F$_{\rm x,2}$ (10$^{-14}$ ergs cm$^{-2}$ s$^{-1}$)    & 6.29 (14.1)            & 2.73 (4.7)            & ...                        \\
log L$_{\rm x}$ (ergs s$^{-1}$)                       & 29.65                  & 29.48                 & 29.41                      \\
log (L$_{\rm x}$/L$_{*}$)\tablenotemark{c}            & $-$5.31                & $-$5.48               & $-$5.55                    \\
\enddata
\tablecomments{
Based on  simultaneous fits of background-subtracted EPIC pn, MOS1, and MOS2 spectra from 
ObsId 0761790101 binned to a 
minimum of 15 counts per bin using XSPEC v12.10.1. Models include an absorption
component (N$_{\rm H}$) modeled in XSPEC with the $wabs$ model and are 
two-temperature variable metallicity thermal plasma (2T APEC), 
2T variable Ne and Fe abundance thermal plasma (2T VAPEC), and differential 
emission measure (DEM) modeled as $c6pvmkl$ in XSPEC.
The tabulated parameters are the same as in Table 3.
Quantities enclosed in curly braces were held fixed during fitting.
Square brackets enclose 1$\sigma$ confidence intervals.
The total X-ray flux (F$_{\rm x}$) and flux contributions of each temperature
component (F$_{\rm x,i}$)  are the  observed (absorbed) values followed
in parentheses by the unabsorbed value in the 0.2 - 6 keV range.
There is no significant observed flux above 6 keV.
X-ray luminosity (L$_{\rm X}$) is the unabsorbed value in the
0.2 - 6 keV range. A distance of 110 pc is assumed ({\em GAIA} DR2).}
\tablenotetext{a}{Element abundances are relative to the solar values of Anders \& Grevesse (1989).}
\tablenotetext{b}{Measurement uncertainties in observed (absorbed) flux for the simultaneous
                  fits of all three EPIC spectra are $\approx$8\% (1$\sigma$).}
\tablenotetext{c}{Assumes L$_{*}$ = 24 L$_{\odot}$ (Vioque et al. 2019).} 
\end{deluxetable}

\begin{deluxetable}{llcccccc}
\tablewidth{0pt}
\tablecaption{Summary of HL Tau and HD 100546 X-ray Properties}
\tablehead{
           \colhead{Name}               &
           \colhead{ObsId}              &
           \colhead{Date}              &
           \colhead{N$_{\rm H}$}        &
           \colhead{kT$_{1}$, kT$_{2}$} &
           \colhead{F$_{x}$}            &
           \colhead{log L$_{x}$}        &
           \colhead{Rate}               \\
           \colhead{}                   &
           \colhead{}                   &
           \colhead{}                   &
           \colhead{(10$^{22}$ cm$^{-2}$)}   &
           \colhead{(keV)}      &
           \colhead{(ergs cm$^{-2}$ s$^{-1}$)}      &
           \colhead{(ergs s$^{-1}$)}     &
           \colhead{(c s$^{-1}$)}        
                                  }
\startdata
HL Tau    & 11016 (CXO)\tablenotemark{b} & 29 Dec 2009 & 2.4  & 3.0, ...  & 4.91e-13 & 30.50 & 0.034$\pm$0.009    \\
HL Tau    & 20906 (CXO)                  & 27 Dec 2017 & 2.6  & 2.4, ...       & 3.57e-13      & 30.45 & 0.027$\pm$0.012  \\
HL Tau    & 18915 (CXO)                  & 06 Jan 2018 & 2.5  & 4.1, ...       & 7.70e-13      & 30.63 & 0.048$\pm$0.021  \\
HD 100546 & 0761790101 (XMM) & 10 Jul 2015 & 0.25 & 0.3, 1.3  [0.6]\tablenotemark{c} & 8.40e-14 & 29.57 & 0.035$\pm$0.005  \\
\enddata
\tablenotetext{a}{
Notes: X-ray flux F$_{x}$ is the observed (absorbed) value (0.3-8 keV).
X-ray luminosity L$_{x}$ is the intrinsic (unabsorbed) value (0.3-8 keV).
The mean count rates for HL Tau are in the 0.3-8 keV range.  
For HD 100546, the mean background-subtracted count rate  is for the pn detector (0.2-4 keV);
the summed background-subtracted rate of the MOS1$+$MOS2 detectors is 0.020$\pm$0.004 c s$^{-1}$ (0.2-4 keV).
Count rate uncertainties are 1$\sigma$.}
\tablenotetext{b}{Archive data}
\tablenotetext{c}{The value in brackets is the value kT$_{wgtd}$ for the 2T (V)APEC fits
weighted by the contribution to the total emission measure (norm) of each temperature component.}

\end{deluxetable}

\section{Discussion}

\subsection{HL Tau X-ray Emission Processes}

{\em Shocks:}~
The plasma temperature kT = 2.4 - 4.1 keV determined from the X-ray 
spectra of HL Tau is too high to be explained by accretion shocks.
The maximum accretion shock temperature for accreting gas impacting the
star is T$_{s}$ = 2.27$\times$10$^{5}$$\mu$($v_{s}$/100 km s$^{-1}$)$^2$~K,
where the mean mass per particle for fully-ionized solar abundance
plasma is $\mu$ = 0.6 (amu) and $v_{s}$ is the shock speed.
To obtain an upper limit on T$_{s}$, we assume
$v_{s}$ equals the free-fall speed 
$v_{ff}$ = ($\xi$2GM$_{*}$/R$_{*}$)$^{1/2}$. Here,
$\xi$ $\equiv$ [1 - (R$_{*}$/$r_{i}$)] $\leq$ 1 for infalling material
originating at a distance $r_{i}$ from the star.
For HL Tau, we assume M$_{*}$ $\approx$ 1.3 M$_{\odot}$ and
R$_{*}$ $\approx$ 2.1 R$_{\odot}$ (White \& Hillenbrand 2004),
so $v_{ff}$ $\approx$ 484 km s$^{-1}$ if the infalling material
comes in from infinity. Thus, T$_{s}$ = 3.3 MK, or kT$_{s}$ = 0.28 keV,
a factor of $\sim$ 10 less than observed. 
Although cool shock-related plasma at temperatures kT $\approx$ 0.3 keV may be present, 
it would remain undetected because of the strong X-ray absorption of 
HL Tau  which masks  emission below $\sim$1 keV (Fig. 3).

The conclusion reached above also applies to X-ray emission
from the shocked jet. The maximum temperature for a strong
shock is 
T$_{s}$ = 1.5$\times$10$^{5}$($v_{s}$/100 km s$^{-1}$)$^2$~K,
where $v_{s}$ is the velocity of the shock relative to
downstream material (Raga et al. 2002).
The mean absolute jet speed of HL Tau is
$v_{jet}$ $\approx$ 250 km s$^{-1}$ (Movsessian et al. 2012)
which gives a maximum predicted shock temperature 
T$_{s}$ = 0.94 MK, or kT$_{s}$ $\approx$ 0.08 keV,  
well below the observed value. 
In addition, the high X-ray absorption and absence of extended
X-ray structure along the optical jet (Fig. 1)  argue against X-ray emission
originating in a jet offset from the star.

{\em Magnetic Processes:}~
The higher X-ray  temperature and X-ray variability of HL Tau 
are more suggestive of emission originating in magnetically 
confined plasma.  Since the X-ray emitting region is not
spatially-resolved, the precise location(s) relative to the star
where the emission originates is not known.  However, based on
modeling of the large slowly-decaying X-ray flare detected by {\em XMM-Newton},
Giardino et al. (2006) concluded that the flaring plasma arises
in one or more coronal loops extending out to several stellar radii.
This does not preclude X-ray emission originating closer to the 
star in non-flaring (``quiescent'') states.

Previous studies have shown that X-ray luminosity is correlated
with stellar mass in T Tauri stars (Preibisch et al. 2005; Telleschi et al. 2007a).
Using {\em XMM-Newton} data for classical TTS (cTTS) detected in the {\em XEST} survey,
Telleschi et al. (2007a) obtained a regression fit using the bisector 
method (which treats the two variables symmetrically) of the form
log L$_{x}$ = 1.98($\pm$0.20)log(M$_{*}$/M$_{\odot}$) $+$ 30.24($\pm$0.06) ergs s$^{-1}$.
Assuming M$_{*}$ = 1.3 M$_{\odot}$ for HL Tau gives 
log L$_{x}$ = 30.47 (30.38-30.55) ergs s$^{-1}$, which agrees well with the
observations (Table 3). Thus, even though HL Tau is a class I protostar,
its L$_{x}$ is not anomalous compared to cTTS of similar mass in Taurus.
However, the mean X-ray absorption and plasma temperature of HL Tau 
from  our spectral fits of three {\em Chandra} observations 
(N$_{\rm H,avg}$ = 2.5 $\times$ 10$^{22}$, kT$_{avg}$ = 3.3 keV) 
are both at the high end of the range compared to cTTS detected in 
the {\em XEST} Taurus sample (Fig. 13 of Telleschi et al. 2007a).
The higher absorption is anticipated for HL Tau since the 
central protostar is still obscured and infalling gas is present
from a remnant envelope,  
whereas cTTS are still accreting but optically-revealed.

\subsection{HD 100546 X-ray Emission Processes}

The soft X-ray spectrum of HD 100546 provides a sharp contrast to the 
harder spectrum of HL Tau. Although the lower X-ray temperature of
HD 100546 does not in itself rule out a late-type companion as the 
X-ray source, no evidence for a close physical stellar companion was
found in {\em HST} images (Grady et al. 2001). We thus consider
intrinsic emission from the Herbig star  by processes that are
capable of producing cool plasma. Such processes include accretion shocks,
wind shocks, current sheets, and soft coronal emission 
possibly sustained by nonsolar mechanisms.

{\em Accretion Shocks}:~
For HD 100546 we assume 
M$_{*}$ = 2.2 M$_{\odot}$ (Pineda et al. 2019) and R$_{*}$ = 1.7 R$_{\odot}$,
where R$_{*}$ is the equivalent blackbody radius based on
L$_{*}$ = 23.4 L$_{\odot}$ and T$_{eff}$ = 9750 K (Vioque et al. 2018).
The predicted accretion luminosity
for a mass accretion rate $\dot{\rm M}_{acc}$ is
L$_{acc}$ = $\xi$G$\dot{M}_{acc}$M$_{*}$/R$_{*}$
where, as above, $\xi$ = [1 - (R$_{*}$/$r_{i}$)].
Estimates of the HD 100546 accretion rate range from
$\dot{M}_{acc}$ $\sim$  few $\times$ 10$^{-9}$ M$_{\odot}$ yr$^{-1}$ (Grady et al. 2005)
to $\dot{M}_{acc}$ $\sim$ 10$^{-7}$ M$_{\odot}$ yr$^{-1}$ (Fairlamb et al. 2015).
For the estimates below, we adopt the mid-range value
$\dot{M}_{acc}$ $\sim$ 10$^{-8\pm1}$ M$_{\odot}$ yr$^{-1}$.

To obtain a realistic estimate of the shock temperature T$_{s}$ and 
accretion lumionsity L$_{acc}$ we must
consider the possibility that the infalling material originates
at a finite distance from the star ($\xi$ $<$ 1).
If the  material originates at the corotation radius $r_{co}$ 
of a Keplerian disk ($r_{i}$ = $r_{co}$), then the HD 100546 stellar 
parameters give $r_{co}$ = 3.2 R$_{*}$ and
$\xi$ = 0.69. Here we have
used $v$sin$i$ = 65 km s$^{-1}$ (Donati et al. 1997) and have assumed
a stellar inclination relative to the line-of-sight $i$ $\approx$ 50$^{\circ}$ 
(Augereau et al. 2001; Deleuil et al. 2004, Pani\"{c} et al. 2014).
The above gives $v_{s}$ $\approx$ $v_{ff}$ = 586 km s$^{-1}$ and 
T$_{s}$ = 4.7 MK (kT$_{s}$ = 0.4 keV). The predicted accretion luminosity is
log L$_{acc}$ = 33($\pm$1) ergs s$^{-1}$, where the 
range reflects the order-of-magnitude uncertainty in $\dot{M}_{acc}$.

We also consider the  possibility of magnetospheric
accretion in which material originating at the inner edge of the 
accretion disk $r_{i}$ is magnetically channeled onto the star. 
The results of K\"{o}nigl (1991) give
$r_{i}$ = $\beta$$\mu_{*}^{4/7}$(2GM$_{*}$)$^{-1/7}$$\dot{M}_{acc}^{-2/7}$
where $\mu_{*}$ = B$_{*}$R$_{*}^{3}$ and we assume $\beta$ = 1 corresponding
to the classical Alfv\'{e}n radius for spherical accretion.
We consider field strengths B$_{*}$ $\approx$ 100 - 300 G as have
been reported for some HAeBe stars (e.g. Hubrig et al. 2004, 2013;
Bagnulo et al. 2015; J\"{a}rvinen et al. 2019).
Taking $\dot{M}_{acc}$ $\sim$ 10$^{-8}$ M$_{\odot}$ yr$^{-1}$
and  B$_{*}$ $\approx$ 100 G gives
$r_{i}$ $\approx$ 1.3 R$_{*}$, 
$v_{s}$ $\approx$ $v_{ff}$ = 341 km s$^{-1}$ (ignoring centrifugal forces), 
T$_{s}$ = 1.6 MK, kT$_{s}$ = 0.14 keV, and log L$_{acc}$ = 32.6 ergs s$^{-1}$.
For a stronger field B$_{*}$ $\approx$ 300 G we obtain
$r_{i}$ $\approx$ 2.5 R$_{*}$, $v_{s}$ $\approx$ = 546 km s$^{-1}$,
T$_{s}$ = 4.0 MK, kT$_{s}$ = 0.35 keV, and log L$_{acc}$ = 33.0 ergs s$^{-1}$.

Based on the above estimates, an accretion shock could account for
the observed X-ray luminosity (log L$_{x}$ = 29.4 - 29.65 ergs s$^{-1}$) 
and would be capable of producing at least some of the 
cooler plasma present in the X-ray spectrum. Also, an attractive
feature of the accretion shock model is that the X-ray
absorption detected in excess of that expected from A$_{\rm v}$ has a
natural explanation. Accretion shock emission originating
at or near the stellar surface could be absorbed by cool accreting
gas, or by the stellar wind, or both. 

To further test 
the relevance of the accretion model, other diagnostics such as 
electron density in the X-ray plasma and X-ray line
widths are needed, based on grating spectra. In general, 
higher densities are expected for
an accretion shock and some line broadening may be present.
{\em XMM-Newton} Reflection Grating 
Spectrometer (RGS) spectra were automatically obtained simultaneously 
with our EPIC data but the grating spectra are noise-dominated
and not suitable for detailed spectral line analysis. 
However, we note that heavily-broadened UV
lines have been detected in HD 100546 with larger line widths than can
be accounted for by thermal effects or stellar rotation
(Deleuil et al. 2004).

{\em Wind Shocks}:~Herbig stars are known to drive winds as
evidenced by P-Cygni type absorption features in emission line
profiles, but mass-loss rates are still rather uncertain.
For HD 100546 we adopt $\dot{M}$ $\sim$ 10$^{-8}$ M$_{\odot}$ yr$^{-1}$
based on the estimate of Wright et al. (2015) and typical radio-derived 
mass-loss rates for similar Herbig stars
such as  AB Aur (G\"{u}del et al. 1989; Skinner et al. 1993; Rodriguez et al. 2014), 
and terminal wind speed $v_{\infty}$ $\sim$ 350 km s$^{-1}$ (Grady et al. 2005).
The kinetic wind luminosity is
L$_{w}$ = (1/2)$\dot{M}$$v_{\infty}^2$, which gives
log L$_{w}$ = 32.6 ergs s$^{-1}$, about three orders
of magnitude greater than L$_{x}$. Thus, there is sufficient
wind kinetic energy to account for the X-ray luminosity,
even at low conversion efficiencies.

The Eddington parameter is
$\Gamma_{e}$ =  ($\sigma_{e}$L$_{*}$)/(4$\pi$cGM$_{*}$),
where $\sigma_{e}$ is the electron scattering coefficient 
per unit mass, c is the speed of light, and
and G the gravitational constant. Adopting
$\sigma_{e}$ $\approx$ 0.3 cm$^2$ g$^{-1}$ as a
typical value for hot stars (Lamers \& Cassinelli 1999),
L$_{*}$ = 23.4 L$_{\odot}$, and M$_{*}$ = 2.2 M$_{\odot}$
gives $\Gamma_{e}$ $\approx$ 3 $\times$ 10$^{-4}$ $\ll$ 1.
Thus, there is insufficient radiation pressure to drive
the wind of HD 100546 and X-ray emission models based
on radiative wind shocks are not applicable.

In the magnetically-confined wind shock (MCWS)
picture,  X-rays are produced in shocks near the magnetic
equator when the magnetically-channeled winds from
each hemisphere collide with one another.
The wind can be confined by a stellar magnetic
field B$_{*}$ if the confinement parameter $\eta$ $>$ 1,
where $\eta$ = (B$_{*}^2$R$_{*}^2$)/$\dot{M}$$v_{\infty}$
(ud-Doula \& Owocki 2002). Adopting
$\dot{M}$ $\sim$ 10$^{-8}$ M$_{\odot}$ yr$^{-1}$
as above and 
$v_{\infty}$ = 350 km s$^{-1}$ gives $\eta$ $>$ 1 if
B$_{*}$ $\gtsimeq$ 100 G.  
However, the results of Babel \& Montmerle (1997) predict a 
maximum wind shock temperature 
T$_{mcws}$ = 1.38 MK (kT$_{mcws}$ = 0.12 keV) if  
$v_{\infty}$ = 350 km s$^{-1}$.
This value is too low to explain the X-ray temperature
determined from HD 100546 spectral fits and the 
presence of spectral lines such as Ne X which form at much higher 
temperatures of $\sim$6 MK.
Thus, unless $v_{\infty}$ is about twice as high as
assumed above, the relevance of the MCWS model seems 
doubtful for HD 100546.

{\em Current Sheets}:~ Usov \& Melrose (1992; hereafter UM92) presented 
a model for X-ray emission
from magnetic early-type stars in which hot gas forms in
a current sheet near the magnetic equator as the wind
drags the field lines outward in a radial direction.
Applying equations 20, 23, 25, and 26 of UM92 to HD 100546 ($\dot{M}$ $\sim$
10$^{-8}$ M$_{\odot}$ yr$^{-1}$, $v_{\infty}$ = 350 km s$^{-1}$,
surface field strength B$_{*}$ $\sim$ 100-300 G)
gives wind absorption 
N$_{\rm H}$ = 3.4$\times$10$^{21}$ cm$^{-2}$,
plasma temperature T = 7.3 MK or kT = 0.63 keV,
and L$_{x}$ $\sim$ 10$^{29}$ - 10$^{30}$ ergs s$^{-1}$. 
These values agree quite well with the HD 100546 observations.
We note, however, that the  model of UM92 assumes
plasma cooling via bremsstrahlung but the X-ray spectrum
of HD 100546 shows several emission lines (Fig. 7), so the
bremsstrahlung cooling assumption is overly simplistic for the case
considered here.

{\em X-ray Corona}:~
In late-type stars with outer convection zones the presence of an 
X-ray emitting corona is linked to a solar-like dynamo. For HAeBe
stars, strong outer convection zones are not expected but thin
convection zones have been discussed as a possible explanation of the
X-ray emission of some well-studied stars such as AB Aur (Telleschi et al. 2007b).
The presence of short-lived outer convection zones due to deuterium burning
in lower mass Herbig  stars (M$_{*}$ $\ltsimeq$ 3.9 M$_{\odot}$) has been 
discussed within the framework of stellar evolution models by Palla \& Stahler (1993).
It has also been suggested by Tout \& Pringle (1995) that young Herbig stars 
might have X-ray emitting coronae as a result of magnetic activity sustained by 
a shear-powered dynamo.

In the shear dynamo model, the X-ray luminosity is predicted to fall off rapidly 
with stellar age $t$ as L$_{x}$($t$) = L$_{x,0}$[1 $+$ ($t/t_{0}$)]$^{-3}$, where the 
reference luminosity L$_{x,0}$ and age $t_{0}$ depend on several factors including
the star's mass, radius, efficiency of magnetic field generation ($\gamma$),
fraction of magnetic flux which heats coronal gas ($\epsilon$), and rotational 
shear $\Delta$$\Omega$ (eqs. [3.15] and [4.4] of Tout \& Pringle 1995). 
In some cases such as
AB Aur, the predicted L$_{x}$ agrees well with observations but
the agreement for other Herbig stars is less satisfactory 
(Skinner et al. 2004; Telleschi et al. 2007b).
Direct comparisons with observations are quite uncertain since
the model predictions depend on several poorly-known parameters
such as $\gamma$ and $\epsilon$.

If we follow the procedure of Skinner et al. (2004) and
adopt fiducial values for the model parameters ($\epsilon$ $\sim$ 10$^{-3}$,
$\gamma$ $\sim$ 3$\times$10$^{-5}$) and the
same stellar parameters as above, along with age
$t$ = 4.8$^{+2.0}_{-1.1}$ Myr for HD 100546 (Pineda et al. 2019),
then the Tout \& Pringle model gives L$_{x,0}$ = 1.92 $\times$ 10$^{31}$ ergs s$^{-1}$,
$t_{0}$ $\approx$ 0.94 Myr, and
log L$_{x}$($t$=4.8 Myr) $\sim$ 28.9 (28.5-29.2) ergs s$^{-1}$.
The range in parentheses reflects the age uncertainty.
The predicted L$_{x}$ is  less than observed,
but the upper limit is only a factor of $\approx$2 below the
observed value (Table 4). That might be considered as 
reasonably good agreement given the uncertainties in the
model and stellar parameters. However, the Tout \& Pringle
model does not predict the X-ray temperature so  
comparison with observational results is quite limited.

{\em Summary}:~Of the mechanisms considered above, the
accretion shock model emerges as the most likely explanation
of the X-ray emission of HD 100546. The current sheet model
can also explain the observed plasma temperature and X-ray
luminosity, but this model assumes bremsstrahlung cooling
whereas the observed X-ray spectrum shows emission lines
that are not anticipated for a pure bremsstrahlung spectrum.
The predicted X-ray luminosity of the shear dynamo soft coronal model
is moderately less than observed, but this discrepancy could
be due to uncertainties in the assumed values of 
efficiency and shear parameters, whose values are not
emprically determined for HD 100546. However, the significant
X-ray absorption by cool gas present in the X-ray spectrum (Sec. 4.3) 
is more difficult to reconcile with coronal models than with models such
as accretion shocks where the X-rays originate at or near the
stellar surface.

\begin{deluxetable}{llcllccccl}
\tablewidth{0pt}
\tablecaption{Herbig Star X-ray Comparison}
\tablehead{
           \colhead{Name}               &
           \colhead{Sp. type}           &
           \colhead{Dist.\tablenotemark{a}}           &          
           \colhead{A$_{\rm v}$}         &
           \colhead{A$_{\rm v}$ ref.\tablenotemark{b}  }         &
           \colhead{N$_{\rm H,xray}$}        &
           \colhead{kT$_{wgtd}$\tablenotemark{c}}        &
           \colhead{log L$_{x}$}        &
           \colhead{log (L$_{x}$/L$_{*}$)}        &
           \colhead{X-ray ref.\tablenotemark{b}}       \\
           \colhead{}                   &
           \colhead{}                   &
           \colhead{(pc)}               &
           \colhead{(mag)}              &
           \colhead{}                   &
           \colhead{(10$^{21}$ cm$^{-2}$)}   &
           \colhead{(keV)}      &
           \colhead{(ergs s$^{-1}$)}     &
           \colhead{}                   &
           \colhead{}
                                  }
\startdata
HD 100546               & B9Vne       & 110 & 0.10 - 1.03   & 4,8,9,10,11   & 2.5   & 0.6 & 29.57 & $-$5.39 & (1)  \\
AB Aur\tablenotemark{d} & A0Ve-B9.5e  & 163 & 0.25 - 0.73   & 4,5,6,7,8,11  & 2.1   & 0.4 & 29.63 & $-$5.63 & (1) \\
AB Aur\tablenotemark{e} & ...         & ... & ...           & ...           & [0.5] & 0.4 & 29.73 & $-$5.53 & (2) \\
HD 163296               & A1Ve        & 102 & 0.19 - 0.30   & 4,6,7,11      & 0.8   & 0.5 & 29.44 & $-$5.48 & (3)
\enddata
\tablenotetext{a}{{\em Gaia} DR2}
\tablenotetext{b}{X-ray refs. (1) this work. (2) Telleschi et al. (2007b)
                  (3) Swartz et al. (2005) 
                  A$_{\rm v}$ refs. (4) van den Ancker et al. (1998) 
                  (5) Roberge et al. (2001) (6) Garcia Lopez et al. (2006) 
                  (7) Hillenbrand et al. (1992) (8) Valenti et al. (2000) 
                  (9) Deleuil et al. (2004) (10) Pineda et al. (2019) (11) Valenti et al. (2003)}
\tablenotetext{c}{The mean X-ray plasma temperature weighted by the contribution to the
                  total emission measure of each component.}
\tablenotetext{d}{Tabulated AB Aur X-ray properties are based on fits of {\em XMM-Newton} archive  data (ObsId 0671960101). See text.}
\tablenotetext{e}{Tabulated AB Aur X-ray properties are based on two-temperature variable abundance fits of 
                  {\em XMM-Newton} ObsId 01011440801 with  N$_{\rm H,xray}$ held fixed (Telleschi et al. 2007b).
                  Their L$_{x}$(0.3-10 keV) value has been adjusted upward to d = 163 pc.}
\end{deluxetable}

\subsection{Similarities Between HD 100546, AB Aur, and HD 163296}

Several X-ray properties of HD 100546, including its soft spectrum,
have been seen in two other Herbig stars, AB Aur and HD 163296 (Table 6).

AB Aur is an extensively studied Herbig Ae star that is often
taken to be the prototype of the class. Its spectral type
of A0Ve - B9.5e is nearly identical to that of HD 100546 (B9Vne).
It is somewhat more distant at a {\em Gaia} DR2 distance of
d = 162.9 (range 161.4 - 164.4) pc. A detailed discussion of its
X-ray properties based on {\em XMM-Newton} EPIC and
Relection Grating Spectrometer (RGS) spectra (ObsId 0101440801; 21-22 Sep 2001)
was given by Telleschi et al. (2007b). Their 2T spectral fits show
a soft X-ray spectrum with characteristic plasma temperature
kT$_{wgtd}$ = 0.45 keV (Table 6), similar to but slightly cooler than HD 100546.

For an additional comparison, we downloaded archive data for
a {\em XMM-Newton} observation of AB Aur obtained on
15-17 Feb 2012 (ObsId 0671960101). The EPIC pn event list was
filtered to remove high background intervals, resulting
in 92,273 s of usable pn exposure. No time filtering was applied
to the MOS data, which had livetimes of 97.45 ks (MOS1) and 97.74 ks (MOS2).
EPIC spectra and associated
response files were extracted using SAS v. 17.0. The
pn spectrum is overlaid on that of HD 100546 in Figure 8.
The spectra are quite similar below $\approx$0.9 keV, but HD 100546
shows stronger emission at higher energies such as the feature
at 1.21 keV that is not visible in the pn spectrum of AB Aur.

Our 2T APEC fits of AB Aur are acceptable and fitting the spectra
of all three EPIC detectors simultaneously gives
N$_{\rm H}$ = 2.1 (1.8-2.4; 1$\sigma$) $\times$ 10$^{21}$ cm$^{-2}$,
kT$_{1}$ = 0.22 (0.20-0.25) keV, kT$_{2}$ = 0.82 (0.79-0.88) keV,
kT$_{wgtd}$ = 0.42 keV, metallicity $Z$ = 0.14 $Z_{\odot}$, absorbed flux
F$_{x,abs}$(0.3-8 keV) = 4.80$\times$10$^{-14}$ ergs cm$^{-2}$ s$^{-1}$,
log L$_{x}$(0.3-8 keV) = 29.63 ergs s$^{-1}$ (d = 163 pc).
If the pn data are excluded and the two MOS spectra are 
fitted simultaneously the kT values are nearly unchanged
but the best-fit absorption decreases slightly to 
N$_{\rm H}$ = 1.9 (1.5-2.5; 1$\sigma$) $\times$ 10$^{21}$ cm$^{-2}$.
The emission measure weighted plasma temperature (kT$_{wgtd}$)
and L$_{x}$ values are nearly the same as determined by
Telleschi et al. (2007b) using earlier AB Aur data acquired in 2001.
However, our fits of the 2012 data suggest a somewhat higher 
N$_{\rm H}$ = 2.0$\pm$0.5 $\times$ 10$^{21}$ cm$^{-2}$
(A$_{\rm V}$ $\approx$ 1.1$\pm$0.3 mag) than their value
N$_{\rm H}$ = 0.5 $\times$ 10$^{21}$ cm$^{-2}$, which was 
based on {\em XEST} results.

Telleschi et al. (2007b) concluded that the X-ray emission of
AB Aur was consistent with predictions of either the current sheet 
model or soft coronal emission, as discussed above 
for HD 100546. They noted that the  observed X-ray temperature
of AB Aur was also compatible with an accretion shock, but electron
densities estimated from O VII triplet line ratios were
about 100 times smaller than predicted by accretion shock models.

The Herbig star HD 163296 (A1Ve) also has similar X-ray properties
to those of HD 100546 (Table 6). Kinematic evidence of one or
more exoplanets orbiting HD 163296 has been reported (Pinte et al. 2018;
Teague et al. 2018).  Analysis of a 19.2 ks {\em Chandra} ACIS-S
observation by Swartz et al. (2005) revealed low absorption
N$_{\rm H}$ = 7.62$\pm$1.85 $\times$ 10$^{20}$ cm$^{-2}$, a soft
spectrum with plasma temperature kT = 0.49$\pm$0.03 keV, and
log L$_{x}$(0.3-3 keV) = 29.44 ergs s$^{-1}$ normalized to the
{\em Gaia} DR2 distance of 101.5 pc. Their fitted  N$_{\rm H}$
value is slightly higher than expected based on A$_{\rm V}$ = 0.25 mag.
No large amplitude X-ray variability was seen but count rate
fluctuations of less than $\approx$20\% were not ruled out.
They considered several  potential emission mechanisms
and concluded that the X-ray properties could be explained
by an accretion shock formed at the stellar surface by
magnetically-channeled infalling gas. However, their analysis
was based on undispersed CCD spectra (as is our
HD 100546 analysis) so precise information
on individual line widths and density-sensitive line flux
ratios was not available. They also pointed out that penetration of
the accretion shock into the stellar photosphere (``buried shock'')
is a potential problem for the accretion shock interpretation.

The remarkable similarity in the X-ray properties of HD 100546, AB Aur,
and HD 163296 (Table 6) leaves little doubt that their X-ray production
involves a common mechanism. It is very unlikely that such a close
similarity would arise if the X-rays were due to close unseen late-type
companions, especially so since no stellar companions have yet been detected.
Consequently, the X-ray emission of these three 
similar Herbig stars is very likely intrinsic but the underlying 
mechanism responsible for the emission is still controversial.
Accretion shock emission is consistent with our results for
HD 100546 and for HD 163296 (Swartz et al. 2005), but
Telleschi et al. (2007b) favored  the current sheet or
soft corona models for AB Aur.

\subsection{X-ray Ionization and Heating of the Disk}

X-rays ionize and heat disk gas, mainly affecting upper disk 
layers above the midplane where most X-ray absorption occurs.
We have estimated the X-ray ionization and heating rates in
the HL Tau and HD 100546 disks using the procedure outlined
below. Our methodology is based on the analytic results of
Shang et al. (2002) and follows the procedures given
in Igea \& Glassgold (1999) and Glassgold et al. (1997a, 1997b, 2004).
The approach is the same as that of our previous analysis of the
LkCa 15 disk (SG17).

As in previous work, the disk is assumed to be aximuthally symmetric 
and dominated by molecular hydrogen with an abundance 
ratio He/H = 0.1 by number. 
Cylindrical coordinates ($r,z$) specify the radial distance $r$ from
the star in the midplane ($z$ = 0)  and height $z$ above the disk midplane. 
The X-ray ionization rate at a given position is (eq. [1] of SG17)

\begin{equation}
\zeta \approx \zeta_{\rm x} \left[{ \frac{r}{R_{\rm x}}} \right]^{-2} \left[{ \frac{kT_{x}}{\epsilon_{ion}}} \right] I_{p}(\tau_{\rm x}, \xi_{0})~~~{\rm s}^{-1}~~{\rm (per~ H~nucleus)}
\end{equation}
where $R_{\rm x}$ is the distance of the X-ray source above (or below) the center of the disk,
$\epsilon_{ion}$ $\approx$ 37 eV  is the energy to create an ion pair,
and the function $I_{p}(\tau_{\rm x}, \xi_{0})$ describes
the X-ray attenuation at optical depth $\tau_{\rm x}(r,z,E)$ and energy $E$.
The attenuation factor $I_{p}(\tau_{\rm x}, \xi_{0})$ is evaluated down to a lower limit
$\xi_{0}$ $\equiv$ $E_{0}$/kT$_{\rm x}$ (eq. [C1] of Shang et al. 2002), where 
$E_{0}$ is the cutoff energy. The value adopted for $E_{0}$ is somewhat arbitrary 
and we use $E_{0}$ = 0.1 keV. Larger values of $E_{0}$ result in slightly flatter
ionization rate versus $\tau_{\rm x}$ curves at small optical depths 
(Glassgold et al. 1997a; Skinner \& G\"{u}del 2013).

The primary X-ray ionization  rate is (eq. [2] of SG17)

\begin{equation}
\zeta_{\rm x} =  \frac{L_{x}\sigma(kT_{x})}{4 \pi R_{x}^2 kT_{x}}
\end{equation}
where $\sigma(kT_{x})$ =  $\sigma(E)$   is the photoelectric X-ray absorption cross-section
per H nucleus evaluated at energy $E = kT_{x}$.
The X-ray cross-section is approximated by a power-law in photon energy of the form
$\sigma(E)$ = $\sigma_{0}$(E/1 keV)$^{-p}$ cm$^{-2}$ where we use
$\sigma_{0}$ = 2.27 $\times$ 10$^{-22}$ cm$^{2}$ and $p$ = 2.485,
as appropriate for solar-abundance disk material.
If heavy elements are depleted then $p$ decreases (Glassgold et al. 1997a).

The  X-ray heating rate per unit volume is proportional to the ionization rate
and is given by 

\begin{equation}
\Gamma_{\rm x} = \zeta n_{\rm H} Q
\end{equation}
where $Q$ is the heating rate per ionization and
$n_{\rm H}$ is the number density of hydrogen nuclei in the disk.
We adopt $Q$ = 17 eV as in previous studies, which 
then gives

\begin{equation}
\Gamma_{\rm x} = 2.72 \times 10^{-11} \zeta  n_{\rm{H}}.
\end{equation}

Tables 7 and 8 give the computed values of $\zeta$ and
$\Gamma_{\rm x}$ for HL Tau and HD 100546, evaluated at
a distance $r$ = 1 au from the star.  The rates are
tabulated for the height $z$ above the disk midplane
at which the X-ray optical depth is unity ($\tau_{x}$ = 1).
X-rays of higher energy will penetrate to greater depths,
all other factors being equal.

For HL Tau, we assume $R_{x}$ = 4 $R_{*}$ to mimic X-ray 
emission from coronal loops. For HD 100546, we use
$R_{x}$ =  $R_{*}$ to represent X-rays originating at or
near the stellar surface as in an accretion shock.
However, the computed values of $\zeta$ and $\Gamma_{\rm x}$ 
far from the star ($r$ $\gg$ R$_{*}$) are not very
sensitive to the assumed value of $R_{x}$ as long as
$R_{x}$ is not larger than a few stellar radii.
Since HD 100546 was fitted with a 2T model (Table 4), separate
rates are given for each temperature component in Table 8,
using kT$_{1}$ = 0.3 keV and kT$_{2}$ = 1.0 keV as
representative temperatures for the 2T fits.

A key parameter is the disk gas surface density 
$\Sigma_{gas}$, which effectively determines the
number density $n_{\rm H}$ in the disk, thus affecting
X-ray absorption and the heating rate. Specifically,
$n_{\rm H}$($r$,$z$) = $\rho$($r$,$z$)/($\mu$m$_{p}$)
where $\rho$ is the mass density, m$_{p}$ is the 
proton mass, and $\mu$ = 1.42 for H-nuclei (Glassgold et al. 2004).
The mass density is
$\rho$($r$,$z$) = $\rho(r,0)$exp[$-$0.5($z$/H($r$))$^2$]
where the midplane density is $\rho(r,0)$ = 0.4[$\Sigma_{gas}(r)$/H($r$)],
and H($r$) is the pressure scale height.
In general, $\Sigma_{gas}$ is not well-determined observationally
close to the star due to limitations on telescope
spatial resolution.  To account for this uncertainty,
we have computed the rates for two different values of 
$\Sigma_{gas}$($r$=1 au) $\equiv$ $\Sigma_{0}$
based on published estimates. For HL Tau, we have
used $\Sigma_{0}$ = 10$^{3}$ and 10$^{2}$ g cm$^{-2}$
as representative values (Pinte et al. 2016; Cridland et al. 2019).
For HD 100546 the surface density is lower and we 
adopt  $\Sigma_{0}$ = 10 and 1 g cm$^{-2}$ for comparison
(Mulders et al. 2013; Pineda et al. 2014). 
As is evident from Tables 7 and 8, for a given 
incident X-ray spectrum the penetration 
depth into the disk decreases as $\Sigma_{0}$ increases.

The values given in Tables 7 and 8 can be scaled to other radii 
but the scaling relations depend on the adopted disk model.
The ionization rate (eq. [1]) scales as
$\zeta(r)$ $\propto$ $r^{-2}$, all other factors
being equal. The heating rate  then scales
as $\Gamma_{\rm x}(r)$ $\propto$ $r^{-2}$$n_{\rm H}(r,z)$.
If power-law dependencies  $\Sigma_{gas}(r)$ $\propto$ $r^{q}$ and 
H($r$) $\propto$ $r^{s}$ are assumed, then
$n_{\rm H}(r,z)$ $\propto$ $r^{q-s}$exp[$-$0.5($z$/H($r$)$^2$].
We have adopted $q$ = $-$1 and $s$ = $+$1.25 and assume 
the disk is in hydrostatic equilibrium with a
radial temperature dependence T($r$) $\propto$ $r^{-0.5}$,
but constant in the $z$ direction.
Multiple variations on the above scaling relations can be found in
the literature. In addition, the above power-law profiles are idealizations
and accounting for structure within the disk requires more
detailed models (e.g. Mulders et al. 2013 for HD 100546).

Despite the differences in disk and X-ray properties,
the main conclusion is similar for both stars. X-ray heating
and ionization are restricted to the uppermost disk layers
at 4-5 scale heights above the midplane. This conclusion
is of particular interest for HD 100546 since the study
of the disk gas by Bruderer et al (2012) concluded that
the gas temperature exceeds the dust temperature
in the inner upper disk by factors of up to 50.
In addition, evidence for warm molecular disk gas 
was reported by Pani\'{c} et al. (2010) on the basis
of CO observations. In both stars, 
$\tau_{x}$($r$=1 au,$z$=0) $\gg$ 1, so
any exoplanets in the midplane at 
$r$ $\geq$ 1 au are well-shielded from X-rays.

\clearpage
\begin{deluxetable}{cccccccc}
\tabletypesize{\footnotesize}
\tablewidth{0pt}
\tablecaption{X-ray Ionization and Heating Rates (HL Tau)}
\tablehead{
           \colhead{r}               &
           \colhead{z/H$_{0}$}              &
           \colhead{E}           &
           \colhead{$\Sigma_{0}$} &
           \colhead{$n_{\rm H}$}         &
           \colhead{N$_{\perp}$}             &
           \colhead{$\zeta$}             &
           \colhead{$\Gamma_{x}$}      \\
           \colhead{(au)}   &
           \colhead{}                 &
           \colhead{(keV)}               &
           \colhead{(g cm$^{-2}$)} &
           \colhead{(cm$^{-3}$)} &
           \colhead{(cm$^{-2}$)} &
           \colhead{(s$^{-1}$)}          &
           \colhead{(ergs s$^{-1}$ cm$^{-3}$)} 
 }
\startdata
  1 & 4.4 &  3.0 & 10$^{3}$ & 1.56e10   & 2.64e21 & 3.18e-11   & 1.35e-11  \\
  1 & 3.8 &  3.0 & 10$^{2}$ & 1.82e10   & 2.64e21 & 3.18e-11   & 1.57e-11  \\
\enddata
\tablecomments{
The ionization rate $\zeta$ (eq. [1]) and heating rate 
$\Gamma_{x}$ (eq. [4]) are computed using
kT = 3 keV and average L$_{x}$ = 3.4 $\times$ 10$^{30}$ ergs s$^{-1}$ (Table 5).
The rates are computed at $r$=1 au for the specified value
$z$/H$_{0}$, the number of scale-heights above the midplane corresponding to $\tau_{x}$ = 1.
Two different assumed values of the gas surface density
$\Sigma_{0}$ $\equiv$ $\Sigma$(r = 1 au) are shown for comparison.
The disk scale height is H$_{0}$ $\equiv$ H(r = 1 au)  = 6.76 $\times$ 10$^{11}$ cm = 0.045 au,
assuming a midplane temperature T($r$=1 au,$z$=0) = 435 K.
The quantity $n_{\rm H}$ is the number density of H-nuclei
at the specified point ($r$,$z$) in the disk.
At the midplane, the adopted disk parameters give
$n_{\rm H}$($r$=1 au,$z$=0) = 2.49 $\times$ 10$^{14}$ cm$^{-3}$
($\Sigma_{0}$ = 10$^{3}$ g cm$^{-2}$), which scales linearly with $\Sigma_{0}$.
The vertically-integrated H-nuclei column density from infinity down 
to the specified height above the midplane is N$_{\perp}$ (eq. [4] of SG17).
The X-ray source is assumed to lie 4 stellar radii above the surface
($R_{x}$/$R_{*}$ = 4) to mimic X-ray emission from coronal loops.
The primary ionization rate (eq. [2]) is
$\zeta_{x}$ = 2.44 $\times$ 10$^{-9}$ s$^{-1}$.
\underline{Scaling Relations}:~See text (Sec. 5.4).
}
\end{deluxetable}
\clearpage

\clearpage
\begin{deluxetable}{cccccccc}
\tabletypesize{\footnotesize}
\tablewidth{0pt}
\tablecaption{X-ray Ionization and Heating Rates (HD 100546)}
\tablehead{
           \colhead{r}               &
           \colhead{z/H$_{0}$}              &
           \colhead{E}           &
           \colhead{$\Sigma_{0}$} &
           \colhead{$n_{\rm H}$}         &
           \colhead{N$_{\perp}$}             &
           \colhead{$\zeta$}             &
           \colhead{$\Gamma_{x}$}      \\
           \colhead{(au)}   &
           \colhead{}                 &
           \colhead{(keV)}               &
           \colhead{(g cm$^{-2}$)} &
           \colhead{(cm$^{-3}$)} &
           \colhead{(cm$^{-2}$)} &
           \colhead{(s$^{-1}$)}          &
           \colhead{(ergs s$^{-1}$ cm$^{-3}$)} 
 }
\startdata
  1 & 4.9 &  0.3 & 10  & 1.47e07   & 1.74e18 & 5.32e-10   & 2.13e-13    \\
  1 & 4.4 &  1.0 & 10  & 1.50e08   & 3.47e19 & 1.43e-11   & 5.83e-14     \\
  1 & 4.4 &  0.3 &  1  & 1.50e07   & 1.74e18 & 5.32e-10   & 2.17e-13     \\
  1 & 3.8 &  1.0 &  1  & 1.76e08   & 3.47e19 & 1.43e-11   & 6.85e-14     \\
\enddata
\tablecomments{
Same as Table 7 except for HD 100546.
Separate values of $\zeta$ and $\Gamma_{x}$ are computed for
cool (E = kT$_{1}$ = 0.3 keV) and warm (E = kT$_{2}$ = 1 keV)
temperature components in 2T thermal spectral fits.
H$_{0}$ $\equiv$ H(r = 1 au)  = 7.03 $\times$ 10$^{11}$ cm = 0.047 au,
assuming  T($r$ = 1 au,$z$=0)) = 865 K.
At the midplane,
$n_{\rm H}$($r$=1 au,$z$=0) = 2.4 $\times$ 10$^{12}$ cm$^{-3}$
if $\Sigma_{0}$ = 10 g cm$^{-2}$, which scales linearly with $\Sigma_{0}$.
The X-ray source is assumed to lie at the stellar surface ($R_{x}$/$R_{*}$ = 1)
to mimic accretion shock emission.
For the cool component: L$_{x,1}$ = 2 $\times$ 10$^{29}$ ergs s$^{-1}$ and
$\zeta_{x}$ = 1.08 $\times$ 10$^{-5}$ s$^{-1}$. For the warm component:
L$_{x,2}$ = 1 $\times$ 10$^{29}$ ergs s$^{-1}$ and 
$\zeta_{x}$ = 8.12 $\times$ 10$^{-8}$ s$^{-1}$.
\underline{Scaling Relations}:~See text (Sec. 5.4).
}
\end{deluxetable}
\clearpage

\section{Summary}

\begin{itemize}

\item We have presented the results of new pointed observations which clarify
      the X-ray properties of the  protostar HL Tau and its TTS neighbor XZ Tau,
      and the Herbig Be star HD 100546. 
      
\item The X-ray emission of  HL Tau is dominated by hot plasma
      (kT $\approx$ 2 - 4 keV) viewed through substantial absorption
      (log N$_{\rm H}$ = 22.4 cm$^{-2}$) and is variable. Its X-ray 
      luminosity based on the new {\em Chandra} observations agrees
      well with that of classical TTS in Taurus of similar 
      mass. However, its X-ray absorption is high (equivalent to
      A$_{\rm V}$ $\approx$ 13 mag), most likely a result of the  
      central star being viewed through an extended remnant envelope of
      infalling gas. No extended soft X-ray emission along the HL Tau
      optical jet near the star was detected. The above properties are 
      typical of magnetically-active late-type coronal sources.
      Modeling of a large X-ray flare detected previously by
      {\em XMM-Newton} supports the idea that the X-rays arise
      in magnetic structures (Giardino et al. 2006).

\item The  classical TTS binary XZ Tau AB located near  HL Tau is a bright
      weakly-absorbed variable X-ray source. No significant X-ray emission 
      along its jet was seen.

\item The X-ray emission of HD 100546 is dominated by cool plasma
      (kT $\leq$ 1 keV) viewed through low but non-negligible absorption 
      (log N$_{\rm H}$ = 21.3 cm$^{-2}$; A$_{\rm v}$ $\approx$ 1 mag). 
      No significant X-ray variability was detected. Its
      X-ray properties are strikingly similar to the Herbig stars
      AB Aur and HD 163296, suggesting that the emission of these
      three stars is intrinsic and due to the same  mechanism,
      and does not originate in unseen late-type companions. Several
      possible X-ray mechanisms were considered for HD 100546 and we 
      conclude that an accretion shock model provides a reasonably good 
      explanation of its X-ray properties. 
      Higher resolution grating spectra capable of measuring fluxes and
      widths of individual emission lines will be needed to 
      further test the accretion shock interpretation and make
      more detailed comparisons with predictions of other models.

\item For both HL Tau and HD 100546, X-ray ionization and heating is
      restricted mainly to upper disk layers well above the midplane.
      Any exoplanets in the midplane at $r$ $\geq$ 1 au are well-shielded
      from X-rays.

\end{itemize} 

\acknowledgments

Support for this work was provided by the National Aeronautics Space Administration (NASA) through
Astrophysics Data Analysis Program (ADAP) award 80NSSC18K0414, and {\em Chandra} award 
number GO6-17133X issued by the 
{\em Chandra} X-ray Center, which is operated by
the Smithsonian Astrophysical Observatory (SAO) for and on behalf of NASA.

\vspace{5mm}
\facilities{{\em Chandra X-ray Observatory}, {\em XMM-Newton}}

\vspace{5mm}
\software{CIAO (Fruscione et al. 2006),
          XSPEC (Arnaud 1996),
          MARX (Davis~ et al. 2012),
          SAS (Gabriel~ et al. 2004)}

\clearpage

\begin{figure}
\figurenum{1}
\includegraphics*[width=9.0cm,angle=0]{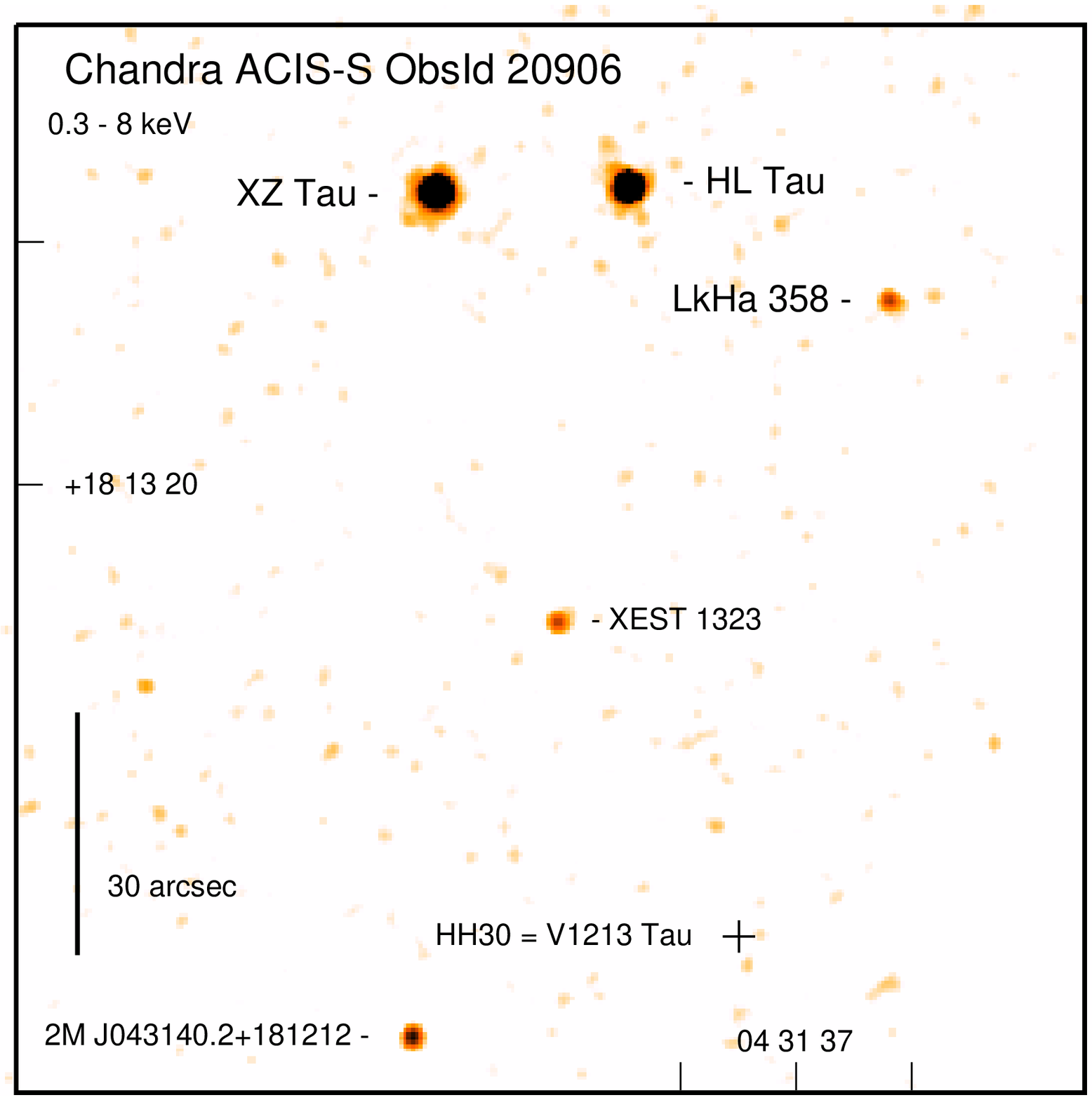} \\
\includegraphics*[width=9.0cm,angle=0]{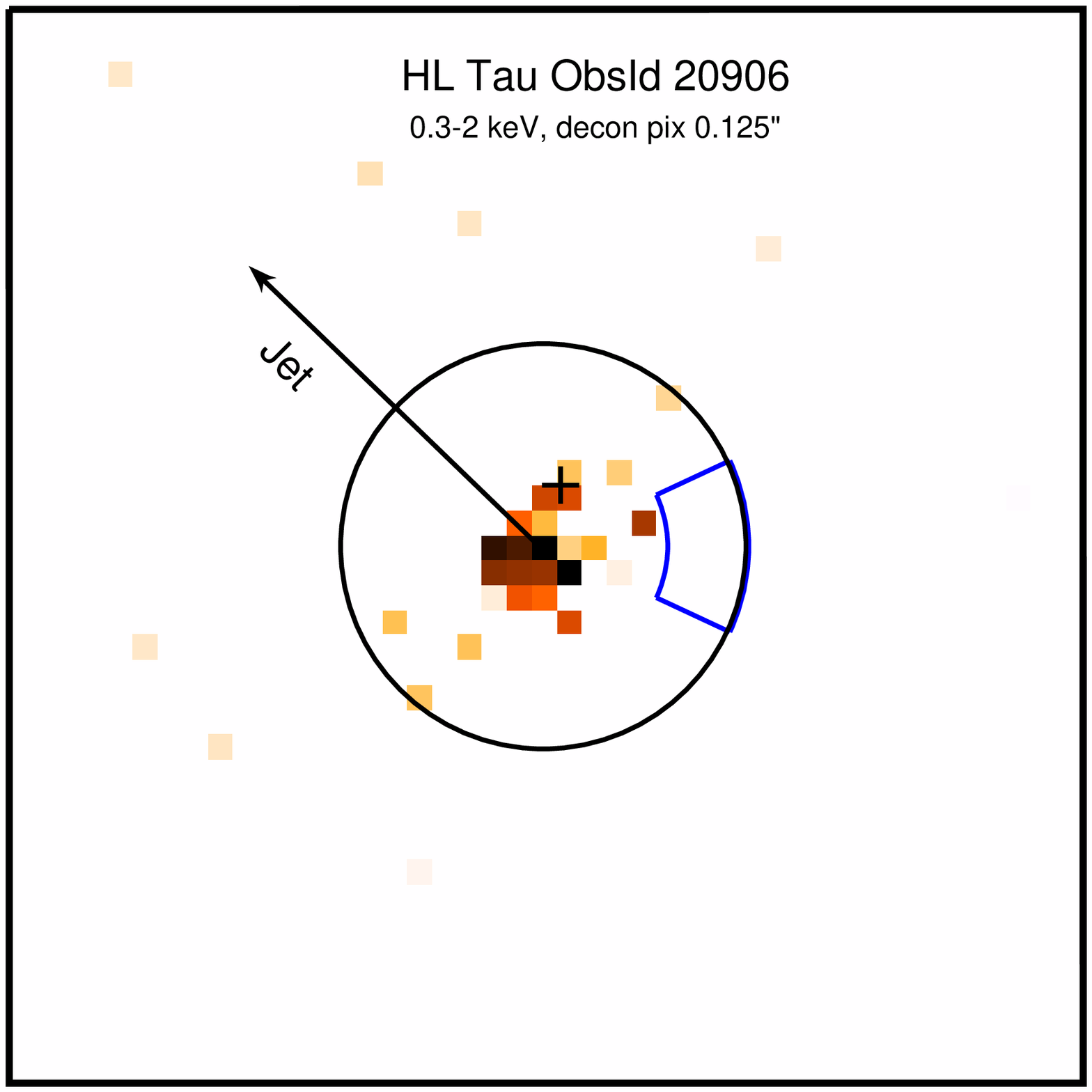}
\caption{Top:~{\em Chandra} ACIS-S image (0.3-8 keV; native 0$''$.492 pixels) of the region near HL Tau
         from ObsId 20906. The image has been lightly Gaussian smoothed using
         a 3-pixel kernel. Young stars in the field are identified.
         V1213 Tau (HH 30) was not detected by {\em Chandra}.
         Bottom:~Deconvolved {\em Chandra} ACIS-S image (0.3-2 keV) of HL Tau
         obtained using the CIAO {\em arestore} algorithm with 50 iterations.
         The image was rebinned to a subpixel size of 0$''$.125 for deconvolution. 
         The cross marks the ALMA position of HL Tau (Brogan et al. 2015).
         The circle centered on the X-ray peak pixel has a radius of 1$''$.
         The sectored region encloses the area that may be affected by the
         ACIS PSF asymmetry (P.A = 270$^{\circ}$$\pm$25$^{\circ}$, 
         0$''$.6 $\leq$ $r$ $\leq$ 1$''$).  
         The arrow marks the direction of the blueshifted optical jet axis along 
         P.A. = 46$^{\circ}$ (Movsessian et al. 2012). There is no significant
         X-ray extension along the jet direction.
         N is up, E is left.
}
\end{figure}

\begin{figure}
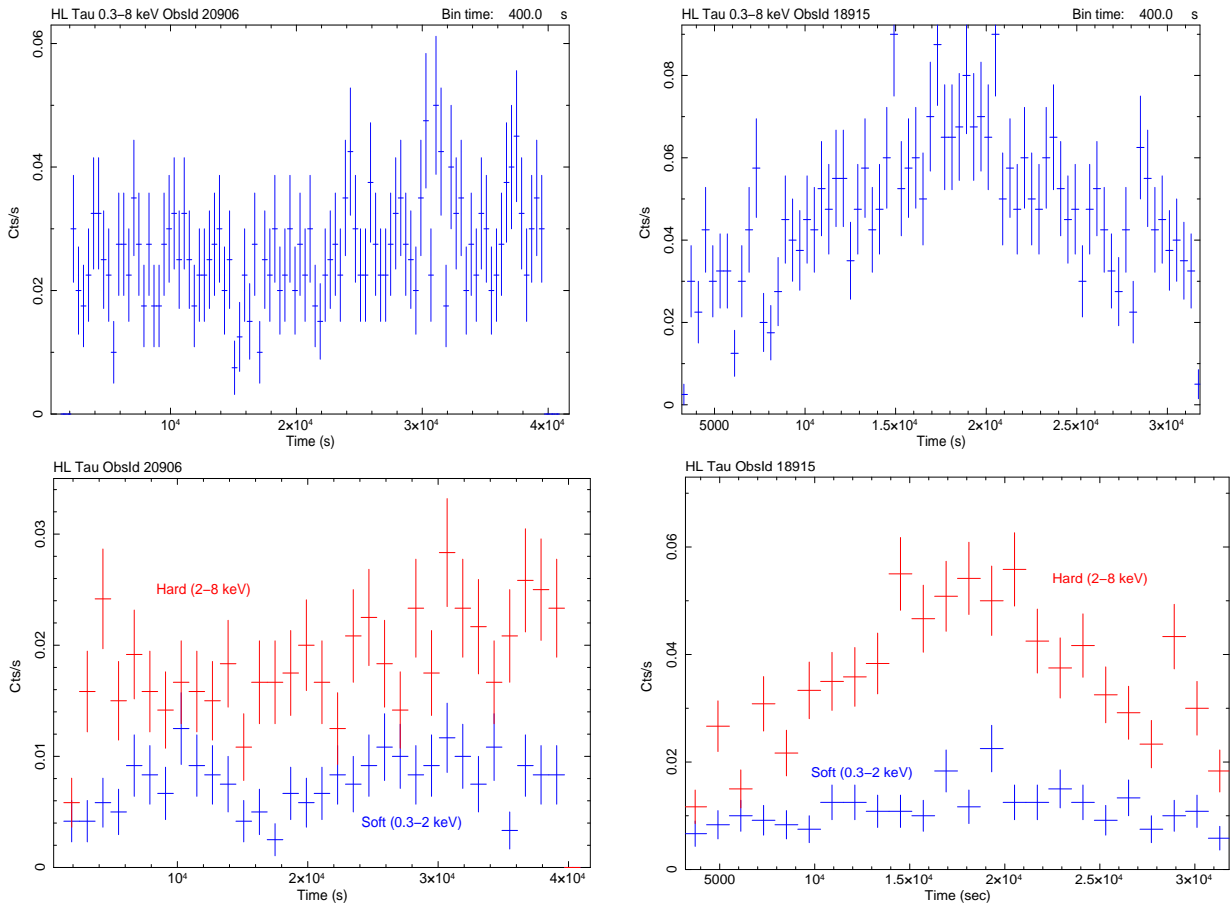

\figurenum{2}
\includegraphics*[width=6.0cm,height=8.26cm,angle=-90]{f2tl.eps}
\includegraphics*[width=6.0cm,height=8.26cm,angle=-90]{f2tr.eps} \\ 
\includegraphics*[width=6.0cm,height=8.26cm,angle=-90]{f2bl.eps}
\includegraphics*[width=6.0cm,height=8.26cm,angle=-90]{f2br.eps}
\caption{Top:~{\em Chandra} ACIS-S X-ray light curves (0.3-8 keV) of HL Tau binned 
at 400 s intervals for ObsId 20906 on 27-28 Dec. 2017 (left) and 18915 on
6 Jan. 2018 (right).
Bottom:~Overlay of soft (0.3-2 keV) and hard (2-8 keV) light curves of HL Tau for ObsId 20906
and 18915, binned at 1200 s intervals.
}
\end{figure}

\clearpage
\begin{figure}
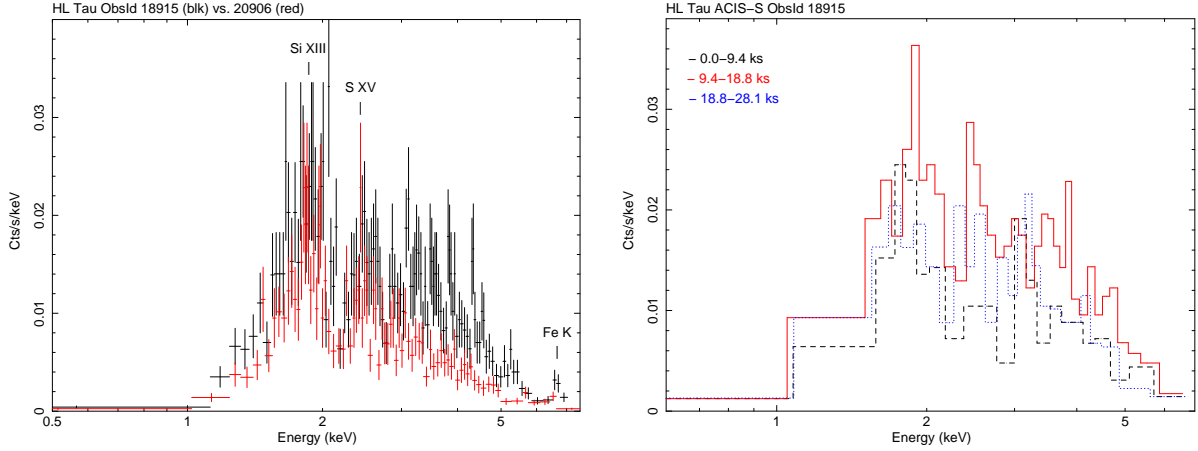

\figurenum{3}
\includegraphics*[width=6.0cm,angle=-90]{f3l.eps} 
\includegraphics*[width=6.0cm,angle=-90]{f3r.eps}
\caption{Left:~{\em Chandra} ACIS-S spectra of HL Tau binned to a minimum of 10
         counts per bin for ObsId 18915 (black) and 20906 (red). 
         A few possible emission
         lines (blends) are identified: Si XIII$r$ (E$_{lab}$ = 1.865 keV),
         S XV (E$_{lab}$ = 2.46 and 2.43 keV), Fe XXV/Fe K complex (E$_{lab}$ = 6.67 keV). 
        Right:~{\em Chandra} time-partitioned ACIS-S spectra of HL Tau binned to a minimum of 15
        counts per bin for ObsId 18915. The exposure was divided into three equal time segments:
        first (black dashed line), middle (red solid), and last (blue dotted). 
        Error bars omitted for clarity.
}
\end{figure}

\begin{figure}
\figurenum{4}
\includegraphics*[width=7.0cm,angle=-90]{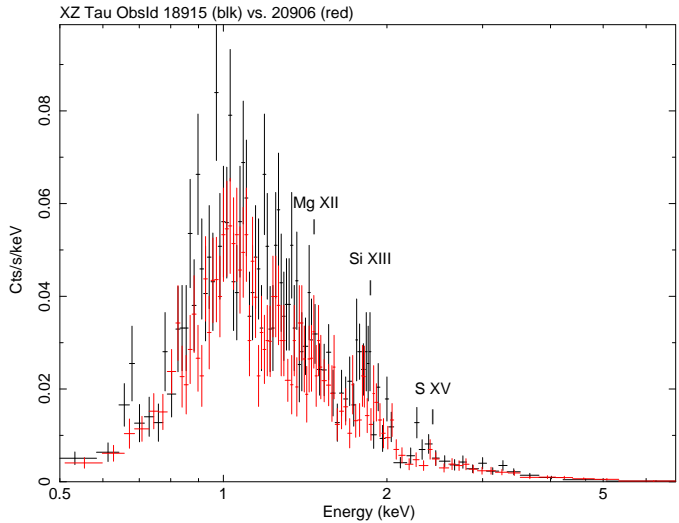} 
\caption{{\em Chandra} ACIS-S spectra of XZ Tau 
         binned to a minimum of 10 counts per bin for ObsId 18915 (black) and
         20906 (red). A few possible emission lines
         are identified. Error bars are 1$\sigma$.
}
\end{figure}

\clearpage                                                                                                                            
\begin{figure}
\figurenum{5}
\includegraphics*[width=6.0cm,angle=0]{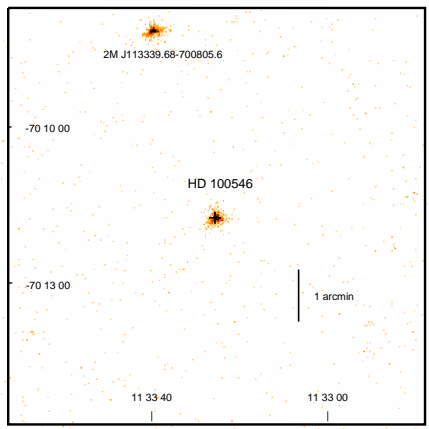} 
\caption{{\em XMM-Newton} binned EPIC MOS2 image of HD 100546 (0.2 - keV).
         Cross at center marks the optical position J113325.44$-$701141.24.
         The X-ray source at top is identified with a K$_{s}$ = 10.54 2MASS source
         classified as a high proper motion star.
         The bounding box dimensions are 8$'$ $\times$ 8$'$.
}
\end{figure}

\begin{figure}
\figurenum{6}
\includegraphics*[width=7.0cm,angle=-90]{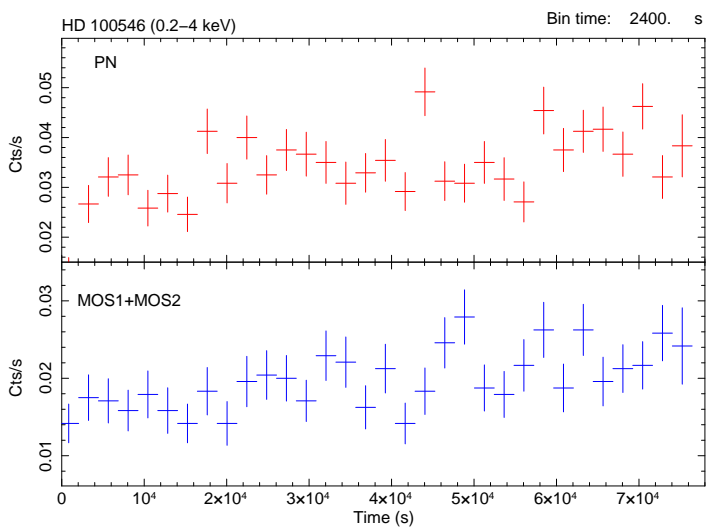} 
\caption{{\em XMM-Newton} EPIC background-subtracted pn and MOS1$+$MOS2 light curves of HD 100546
         in the 0.2-4 keV range, binned at 2400 s intervals. 
         Mean count rates (0.2-4 keV) and standard deviations are 34.7$\pm$5.1 c ks$^{-1}$ (pn) and
         19.7$\pm$3.8 c ks$^{-1}$ (MOS1$+$MOS2). Error bars are 1$\sigma$.
}
\end{figure}

\begin{figure}
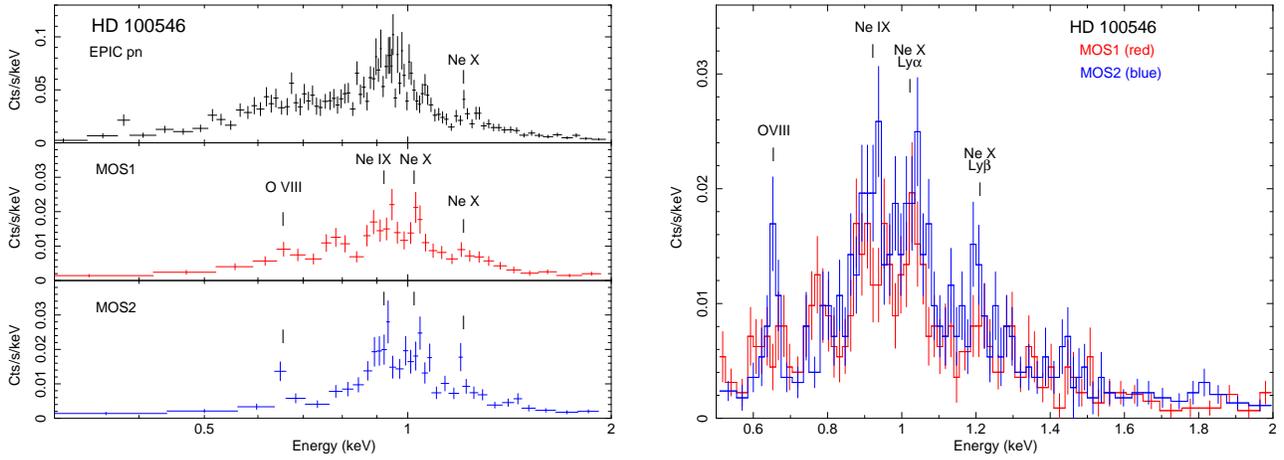

\figurenum{7}
\includegraphics*[width=6.0cm,angle=-90]{f7l.eps}
\includegraphics*[width=6.0cm,angle=-90]{f7r.eps} 
\caption{Left:~{\em XMM-Newton} EPIC background-subtracted spectra of HD 100546
         binned to a minimum of 20 counts per bin. Error bars are 1$\sigma$.
         Right: Overlay of background-subtracted EPIC MOS1 and MOS2 spectra of HD 100546,
         lightly binned to a minimum of 5 counts per bin to bring out the emission lines. 
}
\end{figure}

\begin{figure}
\figurenum{8}
\includegraphics*[width=7.0cm,angle=-90]{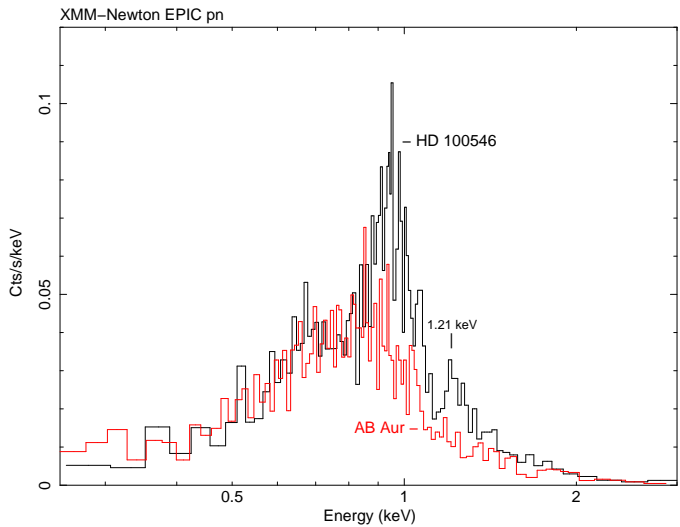} 
\caption{{Background-subtracted \em XMM-Newton} pn histogram spectra of HD 100546 (black) and AB Aur 
         (red; ObsId 0671960101; 2990 net cts) binned to a minimum of 20 counts per bin.
         High background intervals were excluded when extracting the spectra.
         Error bars omitted for clarity. The feature visible at 1.21 keV (Ne X) in HD 100546
         is not present in the AB Aur spectrum.
}
\end{figure}

\end{document}